\documentclass[a4paper, doc]{apa6}
\usepackage{tikz}
\usetikzlibrary{backgrounds}
\usepackage{bbm}
\usepackage{apacite}
\usepackage{natbib}
\usepackage{times}
\usepackage{amssymb,amsmath,amsfonts}
\usepackage[hidelinks]{hyperref}
\usepackage{amsthm}
\usepackage{geometry}
\usepackage{subfig}
\usepackage{graphicx}
\graphicspath{{Figures/}}
\usepackage{placeins}
\usepackage{booktabs}
\usepackage{rotating}
\usepackage[english]{babel}
\usepackage{epstopdf}
\usepackage{multirow}
\usepackage{array}
\usepackage{mathtools}
\usepackage[document]{ragged2e}
\usepackage{threeparttable}
\usepackage{caption}
\usepackage{color}
\usepackage[normalem]{ulem}
\usepackage{float}
\usepackage{comment}
\usepackage{etoolbox}
\patchcmd{\maketitle}
  {\section{\normalfont\normalsize\abstractname}}
  {\section*{\normalfont\normalsize\abstractname}}
  {}{\typeout{Failed to patch abstract.}}

\def\bX{\mathbf{X}}
\def\bZ{\mathbf{Z}}
\def\bz{\mathbf{z}}

\def\bY{\mathbf{Y}}
\def\bW{\mathbf{W}}

\def\E{\mathbb{E}}

\def\bs{\textbf{s}}
\def\bPi{\mathbf{\Pi}}
\def\bomega{\boldsymbol{\omega}}
\def\bGamma{\boldsymbol{\Gamma}}
\def\bSigma{\boldsymbol{\Sigma}}
\def\bB{\boldsymbol{B}}
\def\bgamma{\boldsymbol{\gamma}}
\def\balpha{\boldsymbol{\alpha}}
\def\bPhi{\boldsymbol{\Phi}}
\def\param{\boldsymbol{\theta}}

\def\l1{\ell_\text{STEP 1}}
\def\l2{\ell_\text{STEP 2}}

\newcommand\numberthis{\addtocounter{equation}{1}\tag{\theequation}}

\newtheorem{proposition}{Proposition}[section]

\title{{\Large{\textbf{A two--step estimator for multilevel latent class analysis with covariates}}}}
\shorttitle{two-stage multilevel latent class modeling}

\vspace*{2ex}

\fourauthors{Roberto Di Mari}{Zsuzsa Bakk}{Jennifer Oser}{Jouni Kuha}
 \fouraffiliations{Department of Economics and Business, University of Catania, Italy}{Department of Methodology and Statistics, Leiden University, The Netherlands}{Department of Politics and Government, Ben-Gurion University, Israel}{Department of Statistics, London School of Economics and Political Science, London, UK}
\authornote{Email: roberto.dimari@unict.it\\ Address: Corso Italia 55, 95128, Catania, Italy.}

\abstract{
We propose a two-step estimator for multilevel latent class analysis (LCA) with covariates. 
The measurement model for observed items is estimated in its first step, and in the second step covariates are added in the model, keeping the measurement model parameters fixed. 
We discuss model identification, and derive an Expectation Maximization algorithm for efficient implementation of the estimator. 
By means of an extensive simulation study we show that (i) this approach performs similarly to existing stepwise estimators for multilevel LCA but with much reduced computing time, and (ii) it yields approximately unbiased parameter estimates with a negligible loss of efficiency compared to the one-step estimator. 
The proposal is illustrated with a cross-national analysis of predictors of citizenship norms.
} % 92 words

\begin{document}
\maketitle

\newpage

\section{Introduction}

 Latent class analysis (LCA) is used to create a clustering of units based on a set of observed variables, expressed in terms of an underlying unobserved classification.
 When it is applied to hierarchical (multilevel) data where lower-level units are nested in higher-level ones, the basic latent class model can be extended to account for this data structure. This can be seen as a random coefficients multinomial logistic model (see, for instance \citealp{agresti2000}) for an unobserved categorical variable that is measured by several observed indicators, with a higher-level latent class variable in the role of a categorical random effect \citep{vermunt.03}.  
 Multilevel LCA has become more popular in the social sciences in recent years, for example in educational sciences \citep{fagginger,grilli2022,grilli2016,grilli2011, doi:10.1111/j.2044-8279.2011.02062.x}, economics  \citep{Paccagnella2013}, epidemiology \citep{TOMCZYK2015208, doi:10.1177/002204260603600210, zhang, Lee}, sociology  \citep{costa, morselli1}, and political science  \citep{RUELENS2020102414}.
  In most of these examples, the multilevel LCA  model includes also covariates that are used as predictors of the clustering, and substantive research questions often focus on the coefficients of the covariates.
  
 In estimation of models with covariates, for single-level LCA the current mainstream recommendation is to use \emph{stepwise} methods that separate the estimation of the \emph{measurement model} for the observed indicators from the estimation of the \emph{structural model} for the latent variables given the covariates (see, e.g., \citealp{bakk+kuha18,dimari_bakk_punzo2020,dimari2021,vermunt:10}). 
This is practically convenient because when changes of covariates are made, only the structural model rather than the full model needs to be re-estimated. 
Different structural models can be considered even by different researchers at different times. 
Stepwise estimation can also avoid biases which can arise when all the parameters are instead estimated together in a simultaneous (\emph{one-step}) approach to estimation.
In such cases, misspecifications in one part of the model can cause bias also in the parameter estimates in other parts \citep{bakk+kuha18}. 

In multilevel LCA, the one-step approach is particularly cumbersome because of increased estimation time, especially with multiple covariates possibly defined at different levels.
In that context, there is still need for further research on bias-adjusted efficient stepwise estimators.  
Recently \cite{bakk2021sem} and \cite{jrss2021} proposed a ``two-stage'' estimator for this purpose. The parameters of the measurement model are estimated in its first stage, without including the covariates. 
This is further broken down into three steps. In the first of them, initial estimates of the measurement model are obtained from a single-level LC model, ignoring the multilevel structure.
The latent class probabilities of the multilevel LC model are then estimated, keeping the measurement parameters from the first step fixed. 
Third, to stabilize the estimated measurement model and to account for possible interaction effects, the 
multilevel model is estimated again, now keeping the latent class parameters fixed. 
The estimated measurement parameters from this last step of the first stage are then held fixed in the second stage, where 
the model for the latent classes given covariates is estimated. 

This method has been shown to greatly simplify model construction and interpretation compared to the one-step estimator, with almost identical results if model assumptions are not violated, and with enhanced algorithmic stability and improved speed of convergence. 
In addition, the two-stage estimator exhibits an increased degree of robustness compared to the simultaneous approach in the presence of measurement noninvariance \citep{bakk2021sem}.

%The two-stage approach has some clear benefits over the simultaneous approach (one-step approach) for multilevel LCA models. 

A difficulty in this two-stage technique is deriving an asymptotic covariance matrix that takes into account the multi-step procedure. 
Conditioning on the first-stage estimates as if they were known, even though they are estimates with a sampling distribution, introduces a downward bias in the standard errors, a phenomenon that is well known also in the context of stepwise structural equation models \citep{skrondal2012, oberski}. 
For two-step single level LCA, the standard errors can be corrected in a straightforward way \citep{bakk+kuha18}, but this is more difficult for two-stage LCA due to conditioning on multiple steps. 

The two-stage approach is still in some ways more involved than it needs to be. 
In this paper we show that it is possible to simplify it into a more straightforward \emph{two-step estimator}, still retaining its good performance but with a further reduced computation time. 
This approach is closely motivated by two-step estimation as it is used for single-level LCA. 
In the first step, the full multilevel measurement model is estimated in one go, but without covariates.
In the second step, covariates are included in the model, keeping the measurement model parameters fixed at their estimates from the first step. 

With such a two--step estimator, we contribute to the existing literature in several ways: (1) we establish model identification for the multilevel LC model under standard assumptions, as 
foundation for correct measurement model estimation;  (2) we derive a step-by-step EM algorithm with closed-form formulas to handle the computation of the two-step estimator; and 
(3) we derive the correct asymptotic variance-covariance matrix of the second step estimator of the structural model, drawing on the theory of pseudo maximum likelihood estimation \citep{gong1981pseudo}. 

We evaluate the finite sample properties of our proposal by means of an extensive simulation study. Cross-national data on citizenship norms from the International Association for the Evaluation of Educational Achievement survey are analyzed to illustrate the proposal, and possible extensions are discussed in the conclusions.

\section{The multilevel latent class model with covariates}
Let $\bY_{ij}=(Y_{ij1},\dots,Y_{ijH})'$ be a vector of observed
responses, where $Y_{ijh}$ denotes the response of individual
$i=1,\dots,n_{j}$ in group $j=1,\dots,J$ on the $h$-th categorical
indicator variable (``item''), with $h=1,\dots,H$. 
The data have a hierarchical
(multilevel) structure where the individuals are nested within the
groups. 
In the following, we will also refer to individuals as the
``low-level units'', and groups as the ``high-level units''.
Let $\bY_{j}=(\mathbf{Y}_{1j},\dots,\mathbf{Y}_{n_{j}j})'$ denote the set of responses
for all the low-level units belonging to high-level unit $j$, with $\mathbf{Y}_j$ for different $j$ taken to be independent of each other. 
For simplicity of exposition, we focus below on the case where the items $Y_{ijh}$ are dichotomous, but the idea and methods of two-step estimation proposed here apply in a straightforward way also for polytomous items. 

Let $W_j$ be a categorical latent variable (i.e.\ a \emph{latent class} (LC)
variable) defined at the high level, with possible values $m=1,\dots,M$
and probabilities $P(W_j = m) = \omega_m > 0$, and let
$\boldsymbol{\omega}=(\omega_{1},\dots,\omega_{M})'$. 
Given a realization
of $W_j$, let $X_{ij}$ be a categorical latent variable defined at the
low level, with possible values $t=1,\dots,T$, and conditional
probabilities $P(X_{ij} = t \vert W_j = m) = \pi_{t \vert m} > 0$.
We collect all the $\pi_{t \vert m}$ in the $M \times T$ matrix $\bPi$.
The $X_{ij}$ for the same $j$ are taken to be conditionally independent
given $W_{j}$, so that
\[
P(X_{1j},\dots,X_{n_{j}j})
 = \sum_{m=1}^{M} P(W_{j}=m)
\prod_{i=1}^{n_{j}} P(X_{ij}|W_{j}=m).
\]
This shows that the high-level latent class $W_{j}$ serves as a
categorical random effect which accounts for associations between the
low-level latent classes $X_{ij}$ for different low-level units $i$ within
the same high-level unit $j$.

The items $\mathbf{Y}_{j}$ are treated as observed indicators of the
latent classes.
A multilevel latent class model specifies the probability of observing a
particular response configuration for a high-level unit $j$
as
\begin{eqnarray}
P(\bY_j) &=&
\sum_{m=1}^M P(W_{j}=m) \prod_{i=1}^{n_j} \sum_{t=1}^{T}
P(X_{ij}=t|W_{j}=m)\,P(\mathbf{Y}_{ij}|X_{ij} = t, W_{j}=m)\nonumber \\
&=&\sum_{m=1}^M \omega_m \prod_{i=1}^{n_j} \sum_{t=1}^{T} \pi_{t\vert m}
\prod_{h=1}^{H} P(Y_{ijh}|X_{ij} = t, W_{j}=m),
\label{eq:simplemultileveq}
\end{eqnarray}
where $P(Y_{ijh} \vert X_{ij} = t, W_j = m)$ denotes the conditional
probability mass function of the $h$-th item, given the latent class
variables $X_{ij}$ and $W_{j}$. The second line in this  further assumes that the responses for $Y_{ijh}$ for different items $h$ are conditionally independent given
$(X_{ij},W_{j})$, a standard assumption which we make throughout.

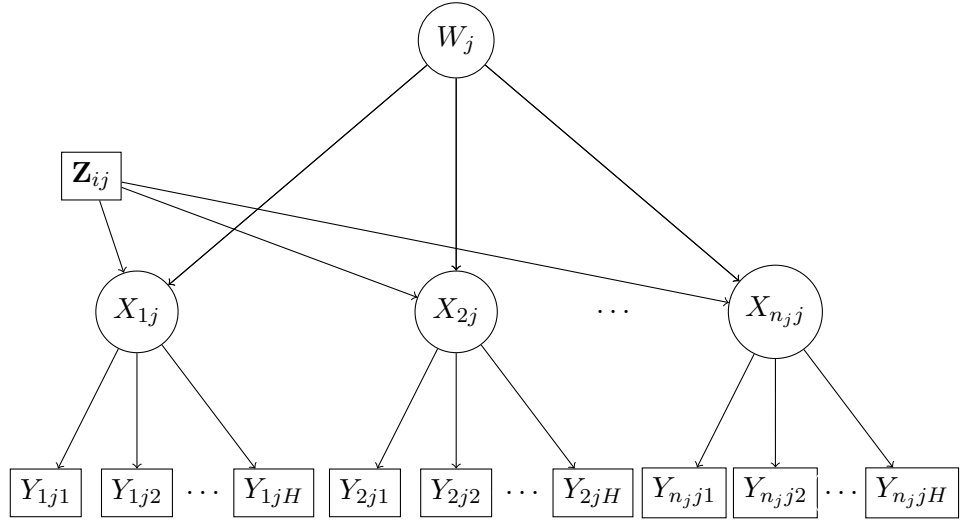
\begin{figure}[!t]
\centering
\begin{tikzpicture}[scale=0.6]
\tikzstyle{every node}=[draw,remember picture] \node[circle] (v1) at (10,8)
{$W_{j}$};
%\node[rectangle] (v11) at (6,8) {$Z_{1j}$};
\node[circle]
(v3) at (3,2) {$X_{1j}$}; \node[circle] (v4) at (10,2)
{$X_{2j}$}; \node[circle] (v5) at (17,2) {$X_{n_{j} j}$};
\node[circle, draw=white] (v101) at (13.5,2) {$\dots$};

\node[rectangle] (v345) at (2,5) {$\bZ_{ij}$};

 \node[rectangle]
(v61) at (1,-2) {$Y_{1j1}$}; \node[rectangle]
(v62) at (3,-2) {$Y_{1j2}$}; \node[circle, draw=white]
(v622) at (4.5,-2) {$\dots$}; \node[rectangle]
(v63) at (6,-2) {$Y_{1jH}$};
\node[rectangle] (v71) at (8,-2) {$Y_{2j1}$}; \node[rectangle] (v72) at (10,-2) {$Y_{2j2}$};
\node[circle, draw=white] (v722) at (11.5,-2) {$\dots$};
\node[rectangle] (v73) at (13,-2) {$Y_{2jH}$};

\node[rectangle] (v81) at (15,-2) {$Y_{n_{j}j1}$};
\node[rectangle] (v82) at (17,-2) {$Y_{n_{j}j2}$};
\node[circle, draw=white] (v822) at (18.5,-2) {$\dots$};
\node[rectangle] (v83) at (20,-2) {$Y_{n_{j}jH}$};

%\draw[->] (v11) -- (v1);
 \draw[->] (v1)
-- (v3); \draw[->] (v1) -- (v4); \draw[->] (v1) -- (v5);
%\draw[->] (v11) -- (v3); \draw[->] (v11) -- (v4);
%\draw[->] (v11) -- (v5);
\draw[->] (v345) -- (v3); \draw[->] (v345) -- (v4);
\draw[->] (v345) -- (v5);
%\draw[->] (v3) -- (v61) node [draw=none, fill=none,midway, above,
%sloped] (TextNode) {fixed}; \draw[->] (v3) -- (v62) node [draw=none,
%fill=none,midway, above, sloped] (TextNode) {fixed}; \draw[->] (v3) --
%(v63) node [draw=none, fill=none,midway, above, sloped] (TextNode)
%{fixed};
%\draw[->] (v4) -- (v71) node [draw=none, fill=none,midway, above, sloped] (TextNode) {fixed}; \draw[->] (v4) -- (v72) node [draw=none, fill=none,midway, above, sloped] (TextNode) {fixed}; \draw[->] (v4) -- (v73) node [draw=none, fill=none,midway, above, sloped] (TextNode) {fixed};
%\draw[->] (v5) -- (v81) node [draw=none, fill=none,midway, above, sloped] (TextNode) {fixed}; \draw[->] (v5) -- (v82) node [draw=none, fill=none,midway, above, sloped] (TextNode) {fixed}; \draw[->] (v5) -- (v83) node [draw=none, fill=none,midway, above, sloped] (TextNode) {fixed};

 \draw[->] (v1)
-- (v3); \draw[->] (v1) -- (v4); \draw[->] (v1) -- (v5); \draw[->] (v3) -- (v61); \draw[->] (v3) -- (v62); \draw[->] (v3) -- (v63);
\draw[->] (v4) -- (v71); \draw[->] (v4) -- (v72); \draw[->] (v4) -- (v73);
\draw[->] (v5) -- (v81); \draw[->] (v5) -- (v82); \draw[->] (v5) -- (v83);

\end{tikzpicture}

\vspace{1ex}
\caption{Graphical representation of a multilevel latent class model which includes a low-level latent class variable $X_{ij}$ nested in a high-level latent class variable $W_j$, and covariates $Z_{ij}$ for $X_{ij}$. Here the response probabilities for items $Y_{ijh}$ depend directly only on $X_{ij}$. 
\label{fig:MLCM}}
\end{figure}

Model \eqref{eq:simplemultileveq} is a general formulation which is equal to an unrestricted multi-group latent class model. Most applications, however, use a more restricted version
which assumes that the item response probabilities do not
depend directly on the high-level latent class $W_{j}$
(\citealp{vermunt.03, Lukociene.10}; 
this model is represented in Figure \ref{fig:MLCM}, if we omit 
the covariates $Z_{ij}$ which will be introduced below). 
We will also make this assumption throughout this paper. 
Model \eqref{eq:simplemultileveq} is also similar to the
multilevel item response model of \cite{gnaldi2016}, but with categorical latent
variables at both levels.
The response probabilities are then given by
\begin{equation}\label{eq:uncondMLC}
P(\bY_j) = \sum_{m=1}^M \omega_m \prod_{i=1}^{n_j} \sum_{t=1}^{T} \pi_{t
\vert m}
\prod_{h=1}^{H} P(Y_{ijh}|X_{ij} = t).
\end{equation}
Therefore, within each high-level latent class $W_{j}$, the model for the items has the form of a standard (single-level) LC model with $X_{ij}$ as the latent class  \citep{mccutcheon:87,goodman:74,hagenaars:90}.
When the items $Y_{ijh}$ are binary with values 0 and 1, we denote
$P(Y_{ijh}=1|X_{ij}=t)=\phi_{h \vert t}$, so that
$P(Y_{ijh}=y_{ijh}|X_{ij}=t)=
\phi_{h|t}^{y_{ijh}}
(1-\phi_{h|t})^{1-y_{ijh}}$, and denote by 
$\bPhi$ the $H\times T$ matrix of all the $\phi_{h|t}$.

It can be shown that the model is identified (in a generic sense, see \citealt{allman2009}), under a
standard set of assumptions:
\begin{proposition}[Identification]\label{prop:ident}
% Under assumpion s i) and ii), provided that $M\leq T$ and $H \geq 3$, the simple multilevel latent class model is identified.
Suppose that
the following conditions hold: (A.1) $\phi_{h\vert t} \neq \phi_{h\vert
s}$ for all $h=1,\dots, H$ and for $t \neq s$; and (A.2) the $M \times T$
matrix $\bPi$ has rank $M$. Then the multilevel LC model
(\ref{eq:uncondMLC}) is identified when $M\leq T$ and $n_j \geq
3$, for all $j=1,\dots,J$.
\end{proposition}

The proof of Proposition \ref{prop:ident} follows the same lines as in
\cite{gassiat2016}, who proved identification of finite state space 
nonparametric hidden Markov models, and applies the results of Theorem 9 of
\cite{allman2009}. 
The fact that all $\phi_{h\vert t}$ are distinct is sufficient for linear independence of the Bernoulli random variables. For $n_j=3$, using the assumption of conditional independence of
low-level units given high-level class $W_{j}$, the distribution of
$(\bY_{1j},\bY_{2j},\bY_{3j})$ factorizes as the product of three terms
$\mu_{ij \vert m} = \sum_{t} \pi_{t \vert m} P(\bY_{ij} \vert X_{ij} =
t)$ for $i=1,2,3$. Assumption (A.2) ensures that $\mu_{1j\vert m}$,
$\mu_{2j\vert m}$ and $\mu_{3j\vert m}$ are linearly independent. Thus
Theorem 9 of \cite{allman2009} applies. 

We make three ancillary comments on Proposition \ref{prop:ident}.
First, for the unrestricted multilevel LC model (\ref{eq:simplemultileveq}), if an assumption analogous to (A.1) holds --- i.e. if all success probabilities of the Bernoulli random variables are distinct --- we can relax (A.2) and prove
identification using \cite{allman2009}'s Theorem 9 (in the related
context of mixture of finite mixtures with Gaussian components, a
similar argument is used by \citealp{dizio2007}). 
Second, for longitudinal and multilevel data, generic identification of the measurement model does not require any condition on the number of items, provided that conditions (A.1) and (A.2) are satisfied.
Third, although we have discussed identification specifically for binary items and Bernoulli conditional distributions, the identification result extends also to polytomous items if we can assume, analogously to (A.1), that all conditional category-class response probabilities are distinct. This guarantees linear independence of the corresponding multinomial random variables. 

%\section{Extending the multilevel LC model to include covariates}
Covariates can be included in the multilevel LC model to predict latent class
membership in both the low and high-level classes. Let
$\mathbf{Z}_{ij}=(1,\mathbf{Z}_{1j}',\mathbf{Z}_{2ij}')'$ be a vector of
$K$ covariates, which can include
high-level  ($\mathbf{Z}_{1j}$)
and low-level 
($\mathbf{Z}_{2ij}$) variables.
For $X_{ij}$ we can consider the multinomial logistic model
\begin{equation}
P(X_{ij}=t \vert W_j = m, \bZ_{ij})=\frac{\exp(\bgamma_{tm}^{\prime} \mathbf{Z}_{ij})}
{1+\sum_{s=2}^{T}\exp(\bgamma_{tm}^{\prime} \mathbf{Z}_{ij})},
\label{eq:logit_covar_ml}
\end{equation}
where $\bgamma_{tm}$ is a $K$-vector of regression coefficients for each
$t=2,\dots,T$ and $m=1,\dots,M$. When only the intercept term is
included, so that $\mathbf{Z}_{ij}=1$, then
$\boldsymbol{\gamma}_{tm}=\log(\pi_{t|m}/\pi_{1|m})$ in the
notation of the model without covariates above.
We denote by $\boldsymbol{\Gamma}$ the $(T-1)M\times K$ matrix of all the parameters in the $\boldsymbol{\gamma}_{tm}$ vectors. 

A model for $W_{j}$ can be specified similarly, now using only
high-level covariates $\mathbf{Z}^{*}_{j}=(1,\mathbf{Z}_{1j}')'$,
as
\begin{equation}
P(W_j = m \vert , \bZ^{*}_{j})=\frac{\exp(\balpha_{m}^{\prime} \bZ^{*}_{j})}{1+\sum_{l=2}^{M}\exp(\balpha_{m}^{\prime} \bZ^{*}_{j})},
\label{eq:logit_covar_ml_Wj}
\end{equation}%
where $\boldsymbol{\alpha}_{m}$ for $m=2,\dots,M$, are regression
coefficients. Although this too is straightforward, for ease of exposition and simplicity of notation we will below not consider models with 
covariates for $W_{j}$, but present the two-step estimator
only for the case where $\mathbf{Z}^{*}_{j}=1$ and thus
$\boldsymbol{\alpha}_{m}=\log(\omega_m/\omega_{1})$. The focus of
interest is then on the model for the low-level (individual-level)
latent class $X_{ij}$, and the high-level (group-level) latent class
$W_{j}$ serves primarily as a random effect which accounts for
intra-group associations between $X_{ij}$. We further assume that the
observed items $\mathbf{Y}_{j}$ are conditionally independent of the covariates
$\mathbf{Z}_{ij}$ given the latent class variables $X_{ij}$.
This means that the measurement of 
$X_{ij}$ by $\mathbf{Y}_{ij}$ is taken to be invariant with respect to the covariates.
With these assumptions, and denoting $\mathbf{Z}_j=(\mathbf{Z}_{1j}',\dots,\mathbf{Z}_{n_{j}j}')'$, the model that we will consider is finally of the form
\begin{equation}\label{eq:condMLC}
P(\bY_j \vert \bZ_{j}) = \sum_{m=1}^M \omega_m \prod_{i=1}^{n_j}
\sum_{t=1}^{T} P(X_{ij} = t \vert W_j = m, \bZ_{ij}) \prod_{h=1}^{H}
P(Y_{ijh}|X_{ij} = t);
\end{equation}
see also a graphical representation of the model in Figure
\ref{fig:MLCM}. 
This model is identified when the corresponding model without covariates is identified, as long as the design matrix of all the $\mathbf{Z}_{ij}$s has full column rank (for an analogous condition for identifiability in the context of single-level latent class models with covariates, see 
\citealt{huang2004} and \citealt{ouyang2022}).

%, if identification of the measurement is established, i.e. of the multilevel latent class model without covariates, if class predictor are introduced, the resulting model is still identified if the \textit{design} matrix $\bZ$ has full column rank.
%See, for an analogous condition for identifiability in the context of single--level latent class models with covariates, \cite{huang2004,ouyang2022}.

\section{Previous methods of estimation}\label{sec:backgroundlit}
\FloatBarrier

We denote the parameters of the model in (\ref{eq:condMLC}) as
$\boldsymbol{\theta}=(\boldsymbol{\theta}_{1}',\boldsymbol{\theta}_{2}')'$
where $\boldsymbol{\theta}_{1}=\text{vec}(\boldsymbol{\Phi})$
are the parameters of the measurement model for the items $\mathbf{Y}_{j}$
and
$\boldsymbol{\theta}_{2}=(
\text{vec}(\boldsymbol{\Gamma})',\boldsymbol{\omega}')'$
the parameters of the structural model the latent
class variables $(X_{ij},W_{j})$ given the covariates $\mathbf{Z}_{ij}$.
Maximum likelihood estimates of these parameters can be obtained by
maximizing the log likelihood $\ell(\boldsymbol{\theta})
=\sum_{j=1}^{J} \log
P(\mathbf{Y}_{j}|\mathbf{Z}_{j})$ with respect to all the parameters
together.
This is the simultaneous or \emph{one-step method} of
estimation for the model. 
It has serious disadvantages, however. 
The full model needs to be re-estimated whenever the covariates in the
structural model are changed, which can be computationally demanding
because of the complexity of such multilevel models. 
Further, because all the parameters are estimated together, misspecification in one part
of the model may destabilize also parameters in other parts of the
model \citep{vermunt:10, asparouhov2012auxiliary}.

Because of the complexity of the one-step approach, in practice the
\emph{classical three-step method} of estimation is more often used. 
In its step 1, model (\ref{eq:uncondMLC}) without covariates is first estimated. 
In step 2, this model is used to assign respondents to the latent classes $X_{ij}$ and
$W_{j}$, conditional on their observed responses $\mathbf{Y}_{j}$; how
this is done for the multilevel LC model is described in detail in
\cite{vermunt.03}.
In step 3 the assigned latent classes are modelled
given covariates, treating the classes now as observed variables.
This is straightforward to do.
However, it, yields biased estimates of the parameters of the structural model, because the assigned classes are potentially misclassified versions of the true latent classes.

Because of this bias in the classical three-step approach, \emph{bias-adjusted stepwise methods} are needed.
One such method for multilevel LC models with covariates is the two-stage estimator proposed by \cite{jrss2021} - see also \cite{bakk2021sem}. It involves the following two stages:
% , and three further steps within the first of them:

\begin{enumerate}
\item[A)] First stage: Unconditional multilevel LC model building (measurement model construction).
\begin{enumerate}
\item[Step 1:] A single-level latent class model is fitted for 
$\mathbf{Y}_{ij}$ given the low-level latent class $X_{ij}$, ignoring
the  multilevel structure of the data. 
This gives an initial estimate of $\boldsymbol{\Phi}$.
\item[Step 2.a:] 
The multilevel model without covariates (equation \ref{eq:uncondMLC}) is estimated, keeping  
$\boldsymbol{\Phi}$ fixed at its estimated value from Step 1.
This gives estimates of $\boldsymbol{\omega}$ and $\boldsymbol{\Pi}$.
\item[Step 2.b:] The two-level model is estimated again, now keeping $\boldsymbol{\omega}$
and $\boldsymbol{\Pi}$ fixed at their estimates from Step 2.a. 
This gives the estimate of $\boldsymbol{\Phi}$ which is taken forward to the second stage. 
\end{enumerate}
\item[B)] Second stage: Inclusion of covariates in the model (structural model construction).
\begin{enumerate}
\item[Step 3:] 
The multilevel model (\ref{eq:condMLC}) with covariates is estimated, keeping the measurement parameters $\boldsymbol{\Phi}$ fixed at their estimates from the first stage.
This gives the two-stage estimates of the structural parameters $\boldsymbol{\theta}_2$.
\end{enumerate}
\end{enumerate}

While effective, the two-stage approach has some shortcomings. 
Although Steps 2.a and 2.b both estimate only part of the measurement model parameters, computationally they do not save much effort because the most challenging part of the estimation (the E-step of the EM algorithm; see below) is required by both steps.
Fixing the response probabilities is also not enough to prevent label switching of the classes from one step to the next in the first stage, since this can simultaneously occur at both the low and high levels.
Finally, estimating the correct form of the second-stage information matrix, which should take variability of the previous steps into account, is difficult due to the sequential  re-updating of the
measurement model. 
These complications make it desirable to look for  more straightforward bias-adjusted stepwise approaches for the multilevel LC model. 
Such a method, the two-step estimator, is described next.

\section{Two--step estimator for the model with covariates}
\FloatBarrier

We propose to amend the two-stage estimator by concentrating all of the measurement modeling into a single step 1, where we estimate the multilevel LC model but without covariates. 
The estimated parameters of the measurement model for the items $\mathbf{Y}_{ij}$ from this step are then taken forward as fixed to step 2, where the structural model for the latent classes given covariates is estimated.
Step 2 is thus the same as the second stage of two-stage estimation, but the three steps of its first stage are here collapsed into the single step 1. 

The two-step estimation procedure for multilevel LC models that is described in this section has been implemented in the R package \texttt{multilevLCA} \citep{lyrvall2023}, which can be downloaded from CRAN. 
The package's routines have been used for the simulations and data analysis in Sections \ref{sec:sim}, and \ref{sec:realdata} of the paper. 

\subsection{Step 1 --- Measurement model}\label{sec:step1}

In the first step, a simple multilevel LC model without covariates
is fitted to the data. Given the data defined above, the log likelihood
for this step is

\begin{equation}\label{eq:ll_LCA}
\ell_{1}=\ell(\bPhi,\bPi,\bomega) = \sum_{j=1}^J \log P(\mathbf{Y}_{j}),
\end{equation}
where $P(\mathbf{Y}_{j})$ is given by \eqref{eq:uncondMLC}. 
This is maximized to find the ML estimate of the parameters of this model.
Direct (numerical) maximization is possible, either with suitable constraints or by adopting well-known logistic re-parametrizations, but it quickly becomes infeasible even for a moderate number of low- and/or high-level classes. 
A more practical alternative to maximize \eqref{eq:ll_LCA} is by means of the Expectation-Maximization (EM) algorithm \citep{EM}, which is what we propose here.

A standard implementation of EM would involve computing $M\times T^{n_j}$ joint posterior probabilities, which is infeasible already with a few low-level units per high-level unit.
Instead, our implementation of the EM algorithm follows closely \cite{vermunt.03}'s \textit{upward--downward} method of computing the joint posteriors of the low- and high-level classes (see also \citealp{vermunt.08}), where the number of joint posterior probabilities to be computed is only a linear function of the number of low-level units per high-level unit.
Here we describe in detail the E and M steps of the algorithm, with the step-by-step implementation, that we use to obtain the estimates in Step 1.

Using standard EM terminology, let us introduce the following augmenting variables:

\begin{equation}
  u_{j,m}  = \begin{cases}
                1, & \mbox{if } W_j = m\\
                0, & \mbox{otherwise}.
              \end{cases}\text{ ,}\quad
  v_{i,j,t,m} = \begin{cases}
                 1, & \mbox{if } X_{ij} = t,\quad W_j = m, \\
                 0, & \mbox{otherwise}.
               \end{cases}
\end{equation}

Defining the \emph{complete-data} sample as\\
$\{\bY_1,\dots,\bY_J,v_{1,1},\dots,u_{j,m},\dots,u_{J,M},v_{1,1,1,1},\dots,v_{i,j,t,m},\dots,v_{n_J,J,T,M}\}$,
the \textit{complete--data log--likelihood} (CDLL) for the first step can be specified as
\begin{align*}\label{eq:cdll_MLCA}
    \ell_{1}^{c} = & \sum_{j=1}^J \sum_{m=1}^M  u_{j,m} \log(\omega_m) + \sum_{j=1}^J \sum_{i=1}^{n_j} \sum_{m=1}^M \sum_{t=1}^{T} v_{i,j,t,m} \log(\pi_{t \vert m}) + \\ &  \sum_{j=1}^J \sum_{i=1}^{n_j} \sum_{m=1}^M \sum_{t=1}^{T} v_{i,j,t,m} \sum_{h=1}^{H} \{ Y_{ijh}\log(\phi_{h \vert t}) + [1-Y_{ijh}]\log(1-\phi_{h \vert t}) \}, \numberthis
\end{align*}
where we have dropped the argument $(\bPhi,\bPi,\bomega)$ from $\ell_{1}^{c}$ for simplicity of notation.

In the E step, the missing data are imputed by conditional expectations given the observed data and
current values for the unknown model parameters.
More specifically, this involves the computation of the following expected CDLL

\begin{align*}\label{eq:Ecdll_MLCA}
   \E\left[ \ell_{1}^{c} \right] = & \sum_{j=1}^J \sum_{m=1}^M  \widehat{u}_{j,m} \log(\omega_m) + \sum_{j=1}^J \sum_{i=1}^{n_j} \sum_{m=1}^M \sum_{t=1}^{T} \widehat{v}_{i,j,t,m} \log(\pi_{t \vert m}) + \\ &  \sum_{j=1}^J \sum_{i=1}^{n_j} \sum_{m=1}^M \sum_{t=1}^{T} \widehat{v}_{i,j,t,m} \sum_{h=1}^{H} \{ Y_{ijh}\log(\phi_{h \vert t}) + [1-Y_{ijh}]\log(1-\phi_{h \vert t}) \} \equiv Q, \numberthis
\end{align*}

where

\begin{equation}\label{eq:Wup}
  \widehat{u}_{j,m} =  \frac{\omega_m \prod_{i=1}^{n_j} \sum_{t=1}^{T} \pi_{t \vert m} \prod_{h=1}^{H} P(Y_{ijh}|X_{ij} = t)}{\sum_{l=1}^M \omega_l \prod_{i=1}^{n_j} \sum_{t=1}^{T} \pi_{t \vert l} \prod_{h=1}^{H} P(Y_{ijh}|X_{ij} = t)}.
\end{equation}

To compute the conditional expectation of $v_{i,j,t,m}$, we use the fact that the joint probability $P(X_{ij} = t, W_j = m \vert \bY_j)$ can be written as $P(W_j = m \vert \bY_j) P(X_{ij} = t \vert W_j,\bY_j)$, where $P(W_j = m \vert \bY_j)$ is already available from \eqref{eq:Wup}.
Note that, given the model assumptions,
\begin{equation}
P(X_{ij} = t \vert W_j,\bY_j) = P(X_{ij} = t \vert W_j,\bY_{ij}),
\end{equation}
which we use to compute the following desired quantity
\begin{align*}\label{eq:XWup}
   \widehat{v}_{i,j,t,m} &=  P(X_{ij} = t, W_j = m \vert \bY_j) \\
    & = P(W_j = m \vert \bY_j) P(X_{ij} = t \vert W_j,\bY_{ij})\\
    & = \widehat{u}_{j,m} \frac{P( X_{ij} = t \vert W_j = m)P(\bY_{ij} \vert X_{ij} = t)}{P(\bY_{ij})} \\
    & = \widehat{u}_{j,m} \frac{\pi_{t \vert m} \prod_{h=1}^{H} P(Y_{ijh}|X_{ij} = t)}{\sum_{s=1}^{T} \pi_{s \vert m} \prod_{h=1}^{H} P(Y_{ijh}|X_{ij} = s)}, \numberthis
\end{align*}
where in the third row we are using the assumption that the joint probability function of the response variables depend on high--level class membership only through low--level class membership.
For the unrestricted multi--group LC model, the expression \eqref{eq:XWup} would be adapted straightforwardly.

In the M step of the algorithm, the expected CDLL \eqref{eq:Ecdll_MLCA} is maximized with respect to the model parameters $(\bPhi,\bPi,\bomega)$ subject to the usual \textit{sum--to--one} constraints on probabilities.
This yields the following closed--form updates

\begin{align}
    \omega_m & = \frac{\sum_{j=1}^J \widehat{u}_{j,m}}{\sum_{j=1}^J \sum_{m=1}^M \widehat{u}_{j,m}}, \label{eq:omegup}\\
    \pi_{t\vert m} & = \frac{\sum_{j=1}^J \sum_{i=1}^{n_j} \widehat{v}_{i,j,t,m}}{\sum_{j=1}^J \sum_{i=1}^{n_j} \sum_{t=1}^{T} \widehat{v}_{i,j,t,m}}, \label{eq:piup}\\
    \phi_{h \vert t} & = \frac{\sum_{j=1}^J \sum_{i=1}^{n_j}\sum_{m=1}^M \widehat{v}_{i,j,t,m} Y_{ijh}}{\sum_{j=1}^J \sum_{i=1}^{n_j} \sum_{m=1}^M \widehat{v}_{i,j,t,m}}. \label{eq:piup}
\end{align}

Starting from initial values for the model parameters, the algorithm iterates between the E- and the M-steps until some convergence criterion is met, e.g.\ until the difference between the log-likelihood values of two subsequent iterations falls below some threshold value.

As for all mixture models, the log-likelihood function can have several local optima and there is no guarantee that the solution found by the EM algorithm is the global optimum \citep{wu1983}.
To better explore the likelihood surface, multiple starting value strategies are typically implemented (among others, see \citealp{biernacki2003,maruotti2021}). 
Beyond doubt, the easiest, and most common approach is to initialize the EM algorithm randomly from several different starting points.
However, even for relatively simpler models, the multiple starting value strategy is often outperformed by more refined techniques \citep{biernacki2003}, .

% \zs{i think this pargraph below about initialization is very relevant and interesting form the programming perspective, but it also brakes the text a bit. i am wondering if we should have this in the main text or as appendix/ endnote. curious to what you think R. btw now this part reads much easier/ clearer than previous version}
For any stepwise estimators, the initialization strategy of earlier steps is particularly relevant
because subsequent steps will be conditional on estimates from previous steps.
In our step 1, we suggest implementing the following hierarchical initialization strategy (for a similar approach in a related context, see for instance \citealp{catania2021,catania2020}):
\begin{itemize}
    \item[(1)] Perform a single--level $K$--modes clustering \citep{huang1997,macqueen1967}, with $K=M$. For each $j=1,\dots,J$
    \begin{itemize}
        \item[-] let $\dot{W}_{ij}$ be the outcome class assignment for unit $i$ in group $j$;
        \item[-] specify $\widetilde{W}_j$ as the most frequent assigned class among the $n_j$ observations belonging to group $j$, and let $\widetilde{W}_{ij} = \widetilde{W}_j$ for all $i=1,\dots,n_j$.
    \end{itemize}
    The relative sizes of the resulting high--level classes are used to initialize $\bomega$.
    The entries of $\bomega$, before being carried over to the actual estimation step, can be sorted in increasing or decreasing order.
    \item[(2)] Fit a single--level $T$--class LC model on the pooled data, ignoring the multilevel structure. 
    Note that the $K$--modes algorithm can be employed herein as well to initialize the single--level LCA.
    The estimated output is organized as follows
    \begin{itemize}
    \item[-] the response probabilities are passed on the EM algorithm as a start for $\bPhi$;
    \item[-] let $\widetilde{X}_{ij}$ be the maximum a posteriori class assignment for unit $i$ in group $j$. 
    Cross--tabulate $\widetilde{\bX}$ and $\widetilde{\bW}$, where $\widetilde{\bX} = (\widetilde{X}_{11},\dots,
    \widetilde{X}_{n_{J}J})^{\prime}$, and $\widetilde{\bW} = (\widetilde{W}_{11},\dots,
    \widetilde{W}_{n_{J}J})^{\prime}$. 
    From the $T \times M$ table of joint counts, compute the conditional (relative) counts of $\widetilde{\bX} \vert \widetilde{\bW}$ to initialize $\bPi$.
    \end{itemize}
    The low--level classes can be re-ordered by letting low--level cluster 1 be the one with the highest average probability to score a ``1'' on all items, cluster 2 the one with the second highest average probability to score a ``1'' on all items, and so on.
\end{itemize}

Note that the suggested rule to re--order low--level classes is only an example of a rule that is often (but not always) useful. 
This is because, if there are many items or some are for rare characteristics, the joint probability of scoring ``1'' on all of them together might be a number so small as to be overwhelmed by sampling error or even by machine imprecision. 
That would effectively bring label switching back again. 
In cases like these, we suggest implementing alternative re--ordering principles.

Running the EM algorithm to convergence from the above starting
values, the solution with the highest log-likelihood \eqref{eq:ll_LCA}
provides us with estimates
$\widehat{\bomega},\widehat{\bPi},\widehat{\bPhi}$.
Of these, $\widehat{\bomega}$ and $\widehat{\bPi}$ are discarded and
$\text{vec}(\widehat{\bPhi})=\widehat{\param}_1$ are retained as the
estimates of the measurement parameters $\boldsymbol{\theta}_{1}$ from
this step 1.

\subsection{Step 2 --- Model for class membership}\label{sec:step2}
\FloatBarrier

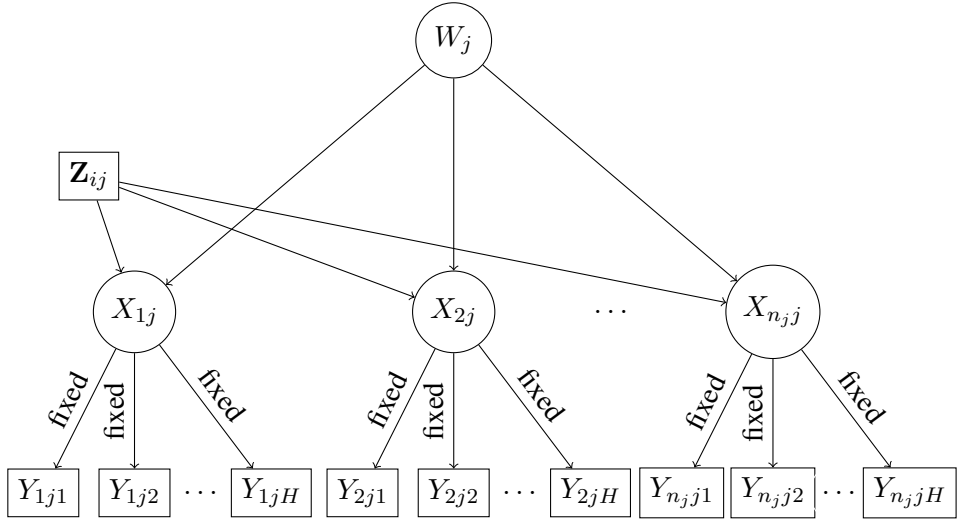
\begin{figure}[!t]
\begin{tikzpicture}[scale=0.6]
\tikzstyle{every node}=[draw,remember picture] \node[circle] (v1) at (10,8)
{$W_{j}$};
%\node[rectangle] (v11) at (6,8) {$Z_{1j}$};
\node[circle]
(v3) at (3,2) {$X_{1j}$}; \node[circle] (v4) at (10,2)
{$X_{2j}$}; \node[circle] (v5) at (17,2) {$X_{n_{j} j}$};
\node[circle, draw=white] (v101) at (13.5,2) {$\dots$};

\node[rectangle] (v345) at (2,5) {$\bZ_{ij}$};

 \node[rectangle]
(v61) at (1,-2) {$Y_{1j1}$}; \node[rectangle]
(v62) at (3,-2) {$Y_{1j2}$}; \node[circle, draw=white]
(v622) at (4.5,-2) {$\dots$}; \node[rectangle]
(v63) at (6,-2) {$Y_{1jH}$};
\node[rectangle] (v71) at (8,-2) {$Y_{2j1}$}; \node[rectangle] (v72) at (10,-2) {$Y_{2j2}$};
\node[circle, draw=white] (v722) at (11.5,-2) {$\dots$};
\node[rectangle] (v73) at (13,-2) {$Y_{2jH}$};

\node[rectangle] (v81) at (15,-2) {$Y_{n_{j}j1}$};
\node[rectangle] (v82) at (17,-2) {$Y_{n_{j}j2}$};
\node[circle, draw=white] (v822) at (18.5,-2) {$\dots$};
\node[rectangle] (v83) at (20,-2) {$Y_{n_{j}jH}$};

%\draw[->] (v11) -- (v1);
 \draw[->] (v1)
-- (v3); \draw[->] (v1) -- (v4); \draw[->] (v1) -- (v5);
%\draw[->] (v11) -- (v3); \draw[->] (v11) -- (v4);
%\draw[->] (v11) -- (v5);
\draw[->] (v345) -- (v3); \draw[->] (v345) -- (v4);
\draw[->] (v345) -- (v5);
 \draw[->] (v3) -- (v61) node [draw=none, fill=none,midway, above, sloped] (TextNode) {fixed}; \draw[->] (v3) -- (v62) node [draw=none, fill=none,midway, above, sloped] (TextNode) {fixed}; \draw[->] (v3) -- (v63) node [draw=none, fill=none,midway, above, sloped] (TextNode) {fixed};
\draw[->] (v4) -- (v71) node [draw=none, fill=none,midway, above, sloped] (TextNode) {fixed}; \draw[->] (v4) -- (v72) node [draw=none, fill=none,midway, above, sloped] (TextNode) {fixed}; \draw[->] (v4) -- (v73) node [draw=none, fill=none,midway, above, sloped] (TextNode) {fixed};
\draw[->] (v5) -- (v81) node [draw=none, fill=none,midway, above, sloped] (TextNode) {fixed}; \draw[->] (v5) -- (v82) node [draw=none, fill=none,midway, above, sloped] (TextNode) {fixed}; \draw[->] (v5) -- (v83) node [draw=none, fill=none,midway, above, sloped] (TextNode) {fixed};
\end{tikzpicture}

\caption{Step 2 of the two-step estimation: Estimating the structural model for
low-level latent classes $X_{ij}$ given covariates $\bZ_{ij}$ and
high-level latent classes $W_j$,
keeping measurement model parameters for items $Y_{ijh}$ fixed at their
estimates from Step 1. \label{fig:step2}}
\end{figure}

In the second step of estimation, the parameters
$\boldsymbol{\theta}_{2}$ of the model for the
latent classes in Equation
\eqref{eq:condMLC} are estimated, keeping the measurement parameters
$\param_1$ fixed at their step-1
estimates $\widehat{\param}_1$ (see Figure \ref{fig:step2}).
These step-2 estimates
are obtained by maximizing
the
pseudo log-likelihood function
\begin{equation}\label{eq:ll_mLCcov}
\ell_{2}(\param_2 \vert \param_1 = \widehat{\param}_1) = \sum_{j=1}^J \log P(\mathbf{Y}_{j} \vert \bZ_{j})
\end{equation}
with respect to $\boldsymbol{\theta}_{2}$. Here $\log P(\mathbf{Y}_{j} \vert \bZ_{j})$ is given by
equation (\ref{eq:condMLC}), except that $\widehat{\param}_1$ are regarded as fixed and known values rather than unknown parameters.
The EM algorithm that we propose for this step
works similarly to the
one that we used for the first step. In particular, under the definition of
the augmenting variables given in Section \ref{sec:step1}, the CDLL is
given by
\begin{align*}\label{eq:cdll_MLCcov}
    \ell_{2}^{c} = & \sum_{j=1}^J \sum_{m=1}^M  u_{j,m} \log(\omega_m) + \sum_{j=1}^J \sum_{i=1}^{n_j} \sum_{m=1}^M \sum_{t=1}^{T} v_{i,j,t,m} \log\left(\frac{\exp(\bgamma_{tm}^{\prime} \mathbf{Z}_{ij})}{1+\sum_{s=2}^{T}\exp(\bgamma_{tm}^{\prime} \mathbf{Z}_{ij})}\right) + \\
    &  \sum_{j=1}^J \sum_{i=1}^{n_j} \sum_{m=1}^M \sum_{t=1}^{T} v_{i,j,t,m} \sum_{h=1}^{H} \{ Y_{ijh}\log(\widehat{\phi}_{h \vert t}) + [1-Y_{ijh}]\log(1-\widehat{\phi}_{h \vert t}) \}, \numberthis
\end{align*}
where we have dropped the argument $(\param_2 \vert \param_1 =
\widehat{\param}_1)$ from $\ell_{2}^{c}$ for ease of notation. Note that the E step is analogous as that described in Section \ref{sec:step1}, except that now the low--level class probabilities conditional on high--level membership depend on covariates.
In the M step the expected CDLL, obtained by substituting the missing
values with expectations computed using analogous formulas as
\eqref{eq:Wup} and \eqref{eq:XWup}, is maximized with respect to
$\param_2$ only. Whereas the update for $\bomega$ is given by
\eqref{eq:omegup}, to derive the update for the regression coefficients
note that $v_{i,j,t,m} = P(X_{ij} = t, W_j = m \vert \bY_j)$ can be
written as the product of $u_{j,m} = P(W_j = m \vert \bY_j)$ and
$q_{i,j,t\vert m} = P(X_{ij} = t \vert W_j,\bY_j)$. Thus, estimates of
$\bGamma$ can be found  solving the
equations
\begin{equation}\label{eq:weightlogis}
    \sum_{j=1}^J \sum_{i=1}^{n_j} \sum_{m=1}^M \sum_{t=1}^{T}
    \widehat{u}_{j,m} \widehat{q}_{i,j,t\vert m} \frac{\partial
    \log\left(P(X_{ij} = t \vert W_j = m, \bZ_{ij})\right)}{\partial
    \text{vec}(\boldsymbol{\Gamma})} = 0,
\end{equation}
which are weighted sums of $M$ equations, each with weights $\widehat{q}_{i,j,t\vert m}$.

%\subsubsection{Starting value strategy and handling label switching}

Stepwise estimation is well known to enhance algorithm stability and
speed of convergence
\citep{bakk+kuha18,bartolucci2015_3,dimari2021,skrondal2012}. However,
class labels in multiple hidden layer models can still be switched, and
keeping the response probabilities fixed cannot prevent it as there are
still $M!$ possible permutations of the high--level class labels. We
handle this issue by initializing $\bomega$ at its estimate from the
first step, and by taking $\log \left( \pi_{t\vert m} / \pi_{1\vert m}
\right )$ to initialize the intercepts $\gamma_{0tm}$, for all
$m=1,\dots,M$ and $t=2,\dots,T$. 
The other elements of $\boldsymbol{\Gamma}$ are initialized at zero. 

\subsection{Selecting the number latent classes}\label{sec:classel}

The description of the two-step estimation procedure above takes the numbers of latent classes at both the lower and higher levels as given.
The selection of these numbers is a separate exercise. 
It is normally carried out without covariates, and the selected numbers of classes are then held fixed when covariates are added. 
This is also in line with general recommendations for LCA with covariates \citep{masyn:dif}.

The selection of the numbers of classes could be considered as a joint exercise of both the high and low levels together, but a generally used recommendation is to use instead a hierarchical
procedure which selects them one at a time \citep{Lukociene.10}.
First, simple LC models are fitted at the lower level and the number of classes for it ($T$) is selected.
Second, this number is held fixed, and multilevel LC models are fitted and compared to select the number of classes at the higher level ($M$).
Third, the selected $M$ is fixed, and model selection for the multilevel model is done again at the lower level, to obtain the final value of $T$.
A still simpler approach would skip the third step \citep{vermunt.03}, but including it allows us to check if the selected number of lower-level classes changes once the within-group associations induced by the high-level classes are allowed for.  

This hierarchical approach can be used with any method of estimating the models.
However, when combined with our two-step estimator, simultaneously selecting the number of classes of the measurement at both levels is also feasible.
Practically, this is possible by leveraging an efficient integration of the above initialization strategy with parallel (multi--core) estimation of all plausible values of $T$ and $M$.

The best candidate values of $M$ and $T$ can be selected with standard information criteria, like AIC or BIC.
For the final choice, we suggest balancing the use information criteria with  the evaluation of low- and high-level class separation, and, perhaps most importantly, the substantive inspection of the candidate model configurations. For a wider discussion on this issue, see, among others, \cite{jrss2021,magidsonvermunt2004}.
In the social sciences, one of the most commonly used measures of class separation is the entropy-based R$^2$ of \cite{magidson81}. The latter can be defined at both lower and higher levels to judge class separation (see \citealp{jrss2021,Lukociene.10}).

\subsection{Statistical properties of the two--step estimator}
Our two--step estimator is an instance of pseudo maximum likelihood
estimation \citep{gong1981pseudo}. Such estimators are consistent and
asymptotically normally distributed under very general regularity
conditions.
The conditions and a proof of consistency can be found in \citet[Sec.
24.2.4]{gourieroux1995}. Let the true parameter vector be
$\param^{\star} = (\param_1^{\star\prime}, \param_2^{\star\prime})^{\prime}$. If the one-step
ML estimator of $\param$ is itself consistent for $\param^{\star}$, in
order to prove consistency of our two-step estimator $\widehat{\param}$
it suffices to show that (1) $\param_1$ and $\param_2$ can vary
independently of each other, and (2) $\widehat{\param}_1$ is consistent
for $\param_1^{\star}$. These conditions are satisfied in our case: (1)
is true by construction of the model, and (2) is
satisfied since $\hat{\boldsymbol{\theta}}_{1}$ from step 1 is a ML
estimate of the measurement model parameters of the multilevel LC model
without covariates, and these parameters are taken to be the same as in
the model with covariates.

%\subsubsection{Asymptotic normality of the two--step estimator}

Let
$\ell(\boldsymbol{\theta}_{1},\boldsymbol{\theta}_{2})$ denote the joint
log-likelihood function for the model, let
$\overline{\bs}_{\param_2}
(\param_1^{\star},\param_2^{\star})$ denote the mean score
$N^{-1} \partial \ell (\param_1,\param_2) / \partial \param_2$
evaluated at $(\param_1^{\star}, \param_2^{\star})$, where $N$
denote the overall sample size, and let
\[
\boldsymbol{{\cal I}}(\boldsymbol{\theta}^{*})=
\begin{bmatrix}
\boldsymbol{{\cal I}}_{11} & \\
\boldsymbol{{\cal I}}_{21} &
\boldsymbol{{\cal I}}_{22}
\end{bmatrix},
\]
be the Fisher information matrix.
In addition, let us suppose that
\[
N^{1/2}
\begin{bmatrix}
\widehat{\param}_1 - \param_1^{\star} & \\
\overline{\bs}_{\param_2} (\param_1^{\star},\param_2^{\star}) &
\end{bmatrix}
\xrightarrow[ ]{d} \text{N}\left( \mathbf{0},
\begin{bmatrix}
\boldsymbol{\Sigma}_{11} & \\
\boldsymbol{\Sigma}_{21} &
\boldsymbol{{\cal I}}_{22}
\end{bmatrix}\right).
\]

Then, using the results of Theorem 2.2 of \cite{gong1981pseudo} (see also \citealp{parke:86}),

\begin{equation}\label{eq:asympnorm}
 N^{1/2} (\widehat{\param}_2 - \param_2^{\star}) \xrightarrow[ ]{d}  \text{N} (\boldsymbol{0},\boldsymbol{V}),
\end{equation}

where $\widehat{\param}_2$ is the proposed two--step estimator and

\begin{equation}
\boldsymbol{V} =
\underbrace{\boldsymbol{{\cal I}}_{22}^{-1}}_{\equiv \boldsymbol{V}_{2}}
+
\underbrace{\boldsymbol{{\cal I}}_{22}^{-1}\,
\boldsymbol{{\cal I}}_{21}\,
\boldsymbol{\Sigma}_{11}\,
\boldsymbol{{\cal I}}_{21}'\,
\boldsymbol{{\cal I}}_{22}^{-1}}_{\equiv \boldsymbol{V}_{1}}.
\label{eq:Vmat}
\end{equation}

Intuitively, $\boldsymbol{V}_{2}$ describes the variability in
$\widehat{\boldsymbol{\theta}}_{2}$ given the step one estimates
$\widehat{\boldsymbol{\theta}}_{1}$, and $\boldsymbol{V}_{1}$ the
additional variability arising from the fact that
$\boldsymbol{\theta}_{1}$ are not known but rather estimated by
$\widehat{\boldsymbol{\theta}}_{1}$ with their own sampling variability.

Let
$\bs_{ij,\param_2}(\widehat{\boldsymbol{\theta}}_{1},\widehat{\boldsymbol{\theta}}_{2})$
be the individual contribution to the score of low--level unit $i$
belonging to high--level group $j$  evaluated at the parameter estimates
of the first and second step respectively. To compute such score we use
the well--known fact that $\partial \ell (\param) / \partial
\param = \partial Q / \partial \param$
\citep{oakes1999}, where $Q= \E \left[ \ell^{c} (\param) \right]$. All
such quantities are available from the above EM algorithm without any
extra effort.  Therefore, $\boldsymbol{{\cal I}}_{22}$ and
$\boldsymbol{{\cal I}}_{21}$ can be estimated respectively as
\begin{equation}
    \widehat{\boldsymbol{{\cal I}}}_{22} = N^{-1} \sum_{j=1}^J \sum_{i=1}^{n_j} \bs_{ij,\param_2}(\widehat{\boldsymbol{\theta}}_{2}) \text{ } \bs_{ij,\param_2}(\widehat{\boldsymbol{\theta}}_{2})^{\prime}
\end{equation}\label{eq:I22es}

and

\begin{equation}
    \widehat{\boldsymbol{{\cal I}}}_{21} = N^{-1} \sum_{j=1}^J
    \sum_{i=1}^{n_j}
    \bs_{ij,\param_2}(\widehat{\boldsymbol{\theta}}_{1},\widehat{\boldsymbol{\theta}}_{2})
    \text{ }
    \bs_{ij,\param_1}(\widehat{\boldsymbol{\theta}}_{1},\widehat{\boldsymbol{\theta}}_{2})^{\prime}.
\end{equation}\label{eq:I12es}

An estimate $\widehat{\mathbf{\Sigma}}_{11}$ can be obtained analogously
by fitting model \eqref{eq:uncondMLC}. We give details on the
derivations of the desired quantities in the appendix.

Note that Equation \eqref{eq:Vmat} shows that there is a loss of
efficiency of the two--step estimator with respect to the simultaneous
ML estimator. This important theoretical and practical aspect with be
investigated in the simulation study --- although we expect this loss to
be rather small as very little information about $\param_2$ should be
contained in $\bY$.

\section{Simulation study}\label{sec:sim}

\subsection{Settings}

We conduct a simulation study to investigate the finite sample properties of the proposed two-step estimator. 
It is compared with the simultaneous (one-step) estimator and the two-stage estimator of \cite{bakk2021sem,jrss2021}. 
One-step estimation is the statistical benchmark, and the two--step estimator's performance is evaluated in terms of its statistical and computational performance relative to this benchmark.  
The target measures that we use for the comparison are the bias,  standard deviations, confidence interval coverage rates, and computation time of the stepwise estimators compared with those of the simultaneous estimator.
We compute both absolute standard deviations, to assess the efficiency of our estimator, as well as relative standard deviations with respect to the one--step method, to investigate potential loss of efficiency with respect to the benchmark.
Class separation and sample size are well-known determinants of the finite--sample behavior of stepwise estimators for LCA \citep{bakk+kuha18,vermunt:10}. 
We considered all combinations of larger and smaller sample sizes, at higher level (30, 50, or 100 higher-level units) and lower level (100 or 500), with a total of 6 sample size conditions. 
Data were generated from a multilevel LC model with 2 high-level classes and 3 low-level classes and with 10 binary indicators and one continuous covariate generated from a standard normal distribution. The random slopes $\gamma_{2 \vert 1}$, and $\gamma_{3 \vert 1}$ were set to -0.25 and -0.25, whereas $\gamma_{2 \vert 2}$, and $\gamma_{3 \vert 2}$ to 0.25 and 0.25, corresponding to a moderate magnitude on the logistic scale.

In multilevel LC models, separation plays a role at both low and high levels \citep{Lukociene.10}. We manipulate low-level class separation by allowing the the response probabilities for the most likely responses to be either 0.7, 0.8 or 0.9, corresponding respectively to low, moderate, and large class separation. We remark that the low class separation condition can be considered as an extreme scenario, in which LCA is hardly carried out in practice.
Nevertheless, we decide to include it as a benchmarking condition. Class profiles are such that the first class has high probability to score 1 on all items, the second class to score 1 on the last five items and 0 on the first 5 items, and the third class is likely to score 0 on all items. At the high level, in the model for $W$, we manipulate class separation by altering the random intercept magnitudes, which are both relatively close to zero in the moderate separation case (-0.85, -1.38 and 0.85, 1.38), and further away from zero in the large separation case (-1.38, -2.07 and 1.38, 2.07). These simulation conditions are in line with previous studies on multilevel LCA \citep{Lukociene.10,park2018}.  

We generated 500 samples for each of the 36 crossed simulation factors of low-level and high-level sample size and low-level and high-level class separation (see Table \ref{tab:simcond}). Data generation and model estimation were carried out in R \citep{R}, with the integration of C++ code for computation efficiency \citep{rcpp}. 

\begin{table}
    \begin{center}
    \begin{tabular}{l c c c c}
    \hline \hline
Condition & LL sample size & HL sample size & LL separation & HL separation \\
\cmidrule{2-5}
1    & 100  & 30  & small  & moderate \\ 
2    & 500  & 30  & small  & moderate \\ 
3    & 100  & 50  & small & moderate \\ 
4    & 500  & 50  & small & moderate \\ 
5    & 100  & 100  & small & moderate \\ 
6    & 500  & 100  & small & moderate \\ 
7    & 100  & 30  & moderate  & moderate \\ 
8    & 500  & 30  & moderate  & moderate \\ 
9    & 100  & 50  & moderate & moderate \\ 
10    & 500  & 50  & moderate & moderate \\ 
11    & 100  & 100  & moderate & moderate \\ 
12    & 500  & 100  & moderate & moderate \\ 
13    & 100  & 30  & large  & moderate \\ 
14    & 500  & 30  & large & moderate \\ 
15    & 100  & 50  & large & moderate \\ 
16    & 500  & 50  & large & moderate \\ 
17    & 100  & 100  & large & moderate \\ 
18    & 500  & 100  & large & moderate \\ 
19    & 100  & 30  & small & large \\ 
20    & 500  & 30  & small & large \\ 
21    & 100  & 50  & small & large \\ 
22    & 500  & 50  & small & large \\ 
23    & 100  & 100  & small & large \\ 
24    & 500  & 100  & small & large \\ 
25    & 100  & 30  & moderate & large \\ 
26    & 500  & 30  & moderate & large \\ 
27    & 100  & 50  & moderate & large \\ 
28    & 500  & 50  & moderate & large \\ 
29    & 100  & 100  & moderate & large \\ 
30    & 500  & 100  & moderate & large \\ 
31    & 100  & 30  & large & large \\ 
32    & 500  & 30  & large & large \\ 
33    & 100  & 50  & large & large \\ 
34    & 500  & 50  & large & large \\ 
35    & 100  & 100  & large & large \\ 
36    & 500  & 100  & large & large \\  

\hline
\hline
    \end{tabular}
    \end{center}
    \caption{24 simulation conditions. LL stands for Low--Level, HL stands for High--Level.}
    \label{tab:simcond}
\end{table}

\FloatBarrier
\subsection{Results}

\begin{figure}[!t]
\centering
\subfloat[One--step estimator \label{fig:bias_onestep}]{
\centering
\includegraphics[width=.6\textwidth]{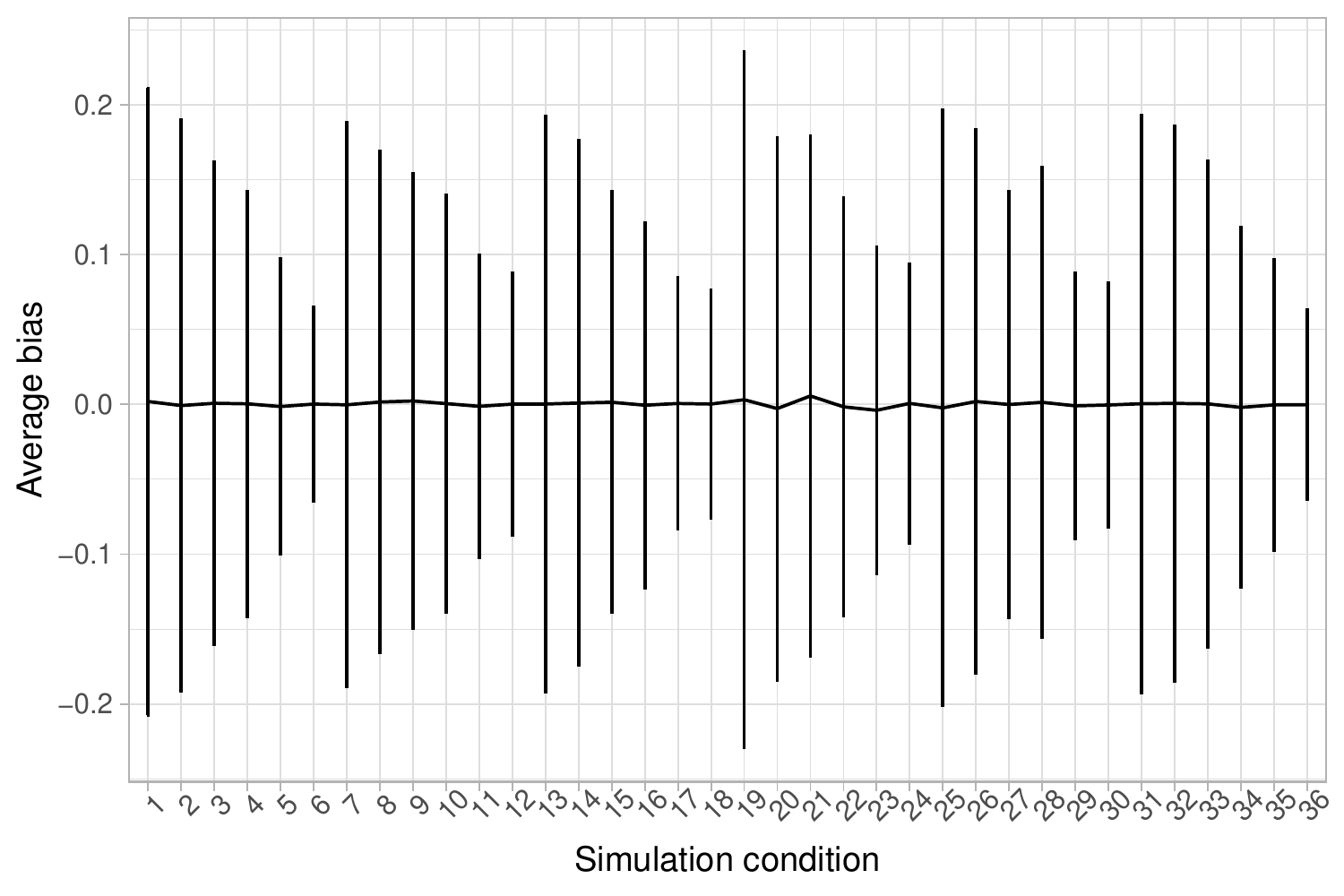}
} \\
\subfloat[Two--stage estimator \label{fig:bias_twostage}]{
\centering
\includegraphics[width=.6\textwidth]{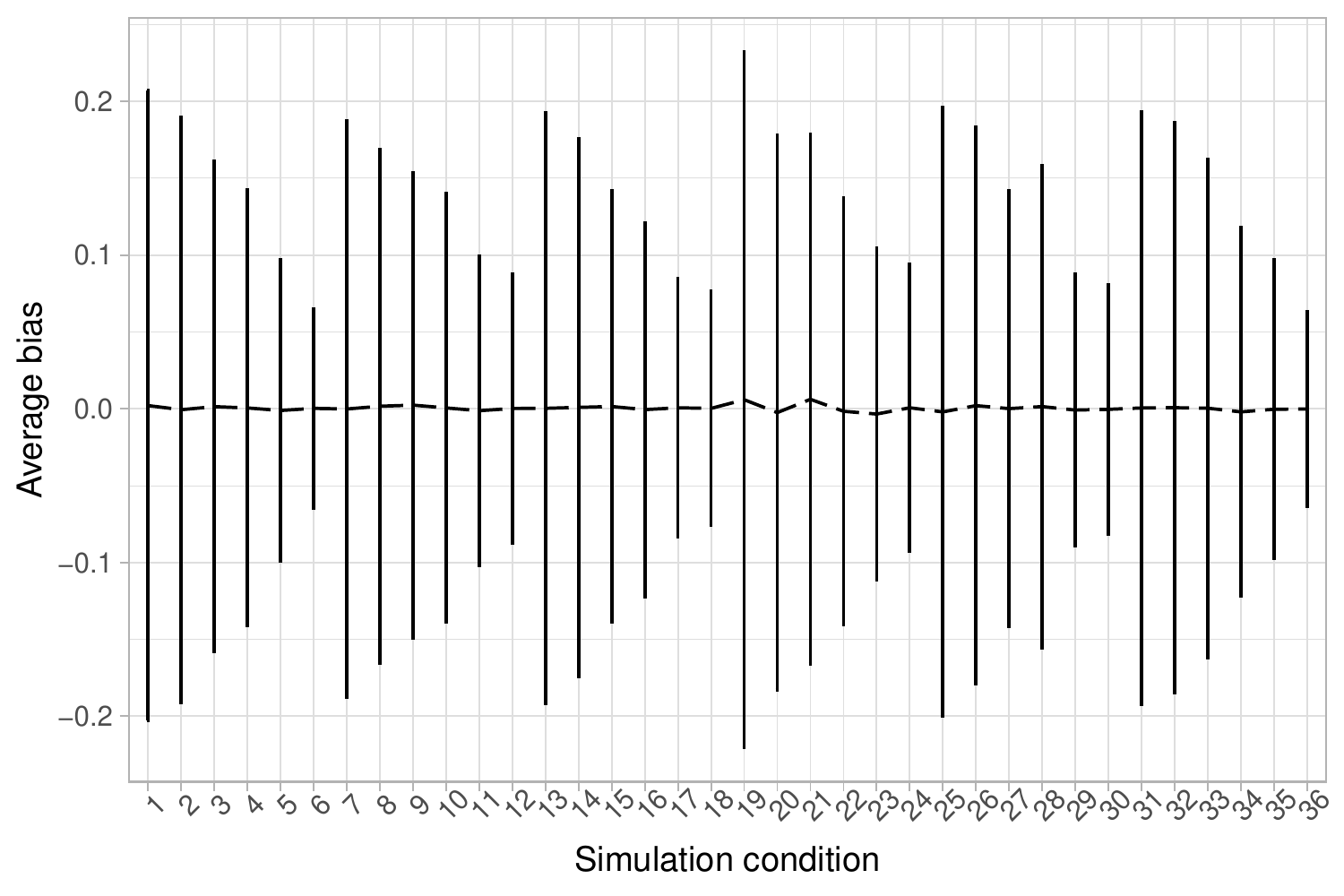}
}\\
\subfloat[Two--step estimator \label{fig:bias_twostep}]{
\centering
\includegraphics[width=.6\textwidth]{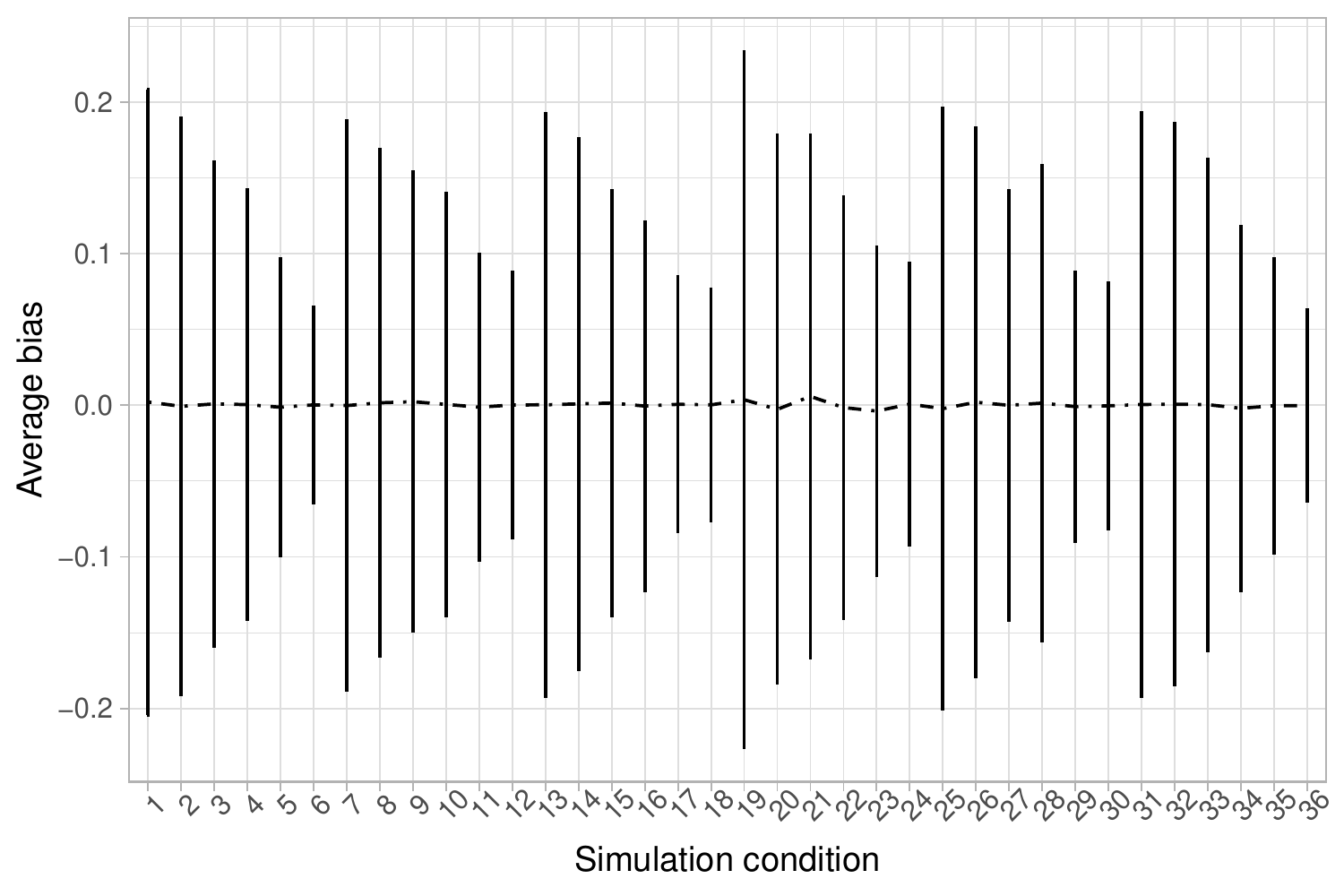}
}

\caption{Line graphs of estimated bias for the one--step, two--step, and two--stage estimators, for the 36 simulation conditions, averaged over the 500 replicates. Error bars are based on mean bias +/- Monte Carlo standard deviations. \label{fig:bias}}
\end{figure}

\begin{figure}
    \centering
    \includegraphics{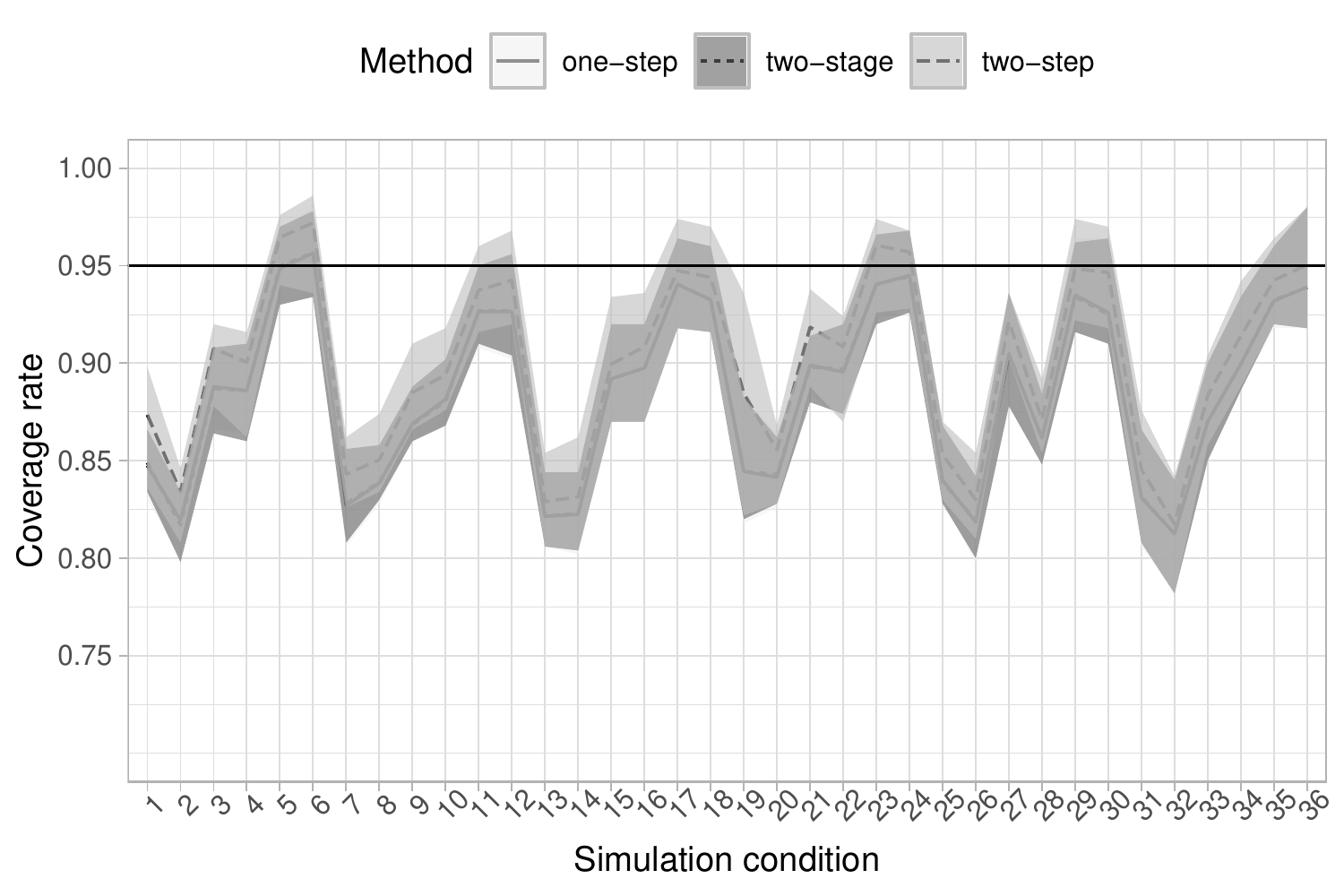}
    \caption{Observed coverage rates of 95\% confidence intervals, averaged over covariate effects, for the one--step, two--stage and two--step estimators for the 36 simulation condition, averaged over the 500 replicates. Lower and higher confidence values reported in the confidence bars, based on the minimum and maximum coverages of the confidence intervals for each covariate effect.}
    \label{fig:CVG}
\end{figure}

All estimators show very similar values for bias (see Figures \ref{fig:bias_onestep}--\ref{fig:bias_twostage}), and both two--stage and two--step estimators have nearly identical results compared to the simultaneous estimator. 
Relative efficiency with respect to the simultaneous estimator (Table \ref{tab:res_sesd}, in the appendix) is, in all conditions, approximately one for both stepwise estimators, with the two-stage estimator doing very slightly worse only in one condition. 
Confidence interval coverages (Figure \ref{fig:CVG}) are mostly very similar between the three estimators. We observe some undercoverage for all methods in the low--separation and small high--level sample size conditions.
This may be due to the fact that expected information matrices are used to estimate the asymptotic variance covariance matrix, rather than the observed ones, and the contributions to the score are computed on high level units, and to the overlap between classes.  

\begin{figure}
    \centering
    \includegraphics{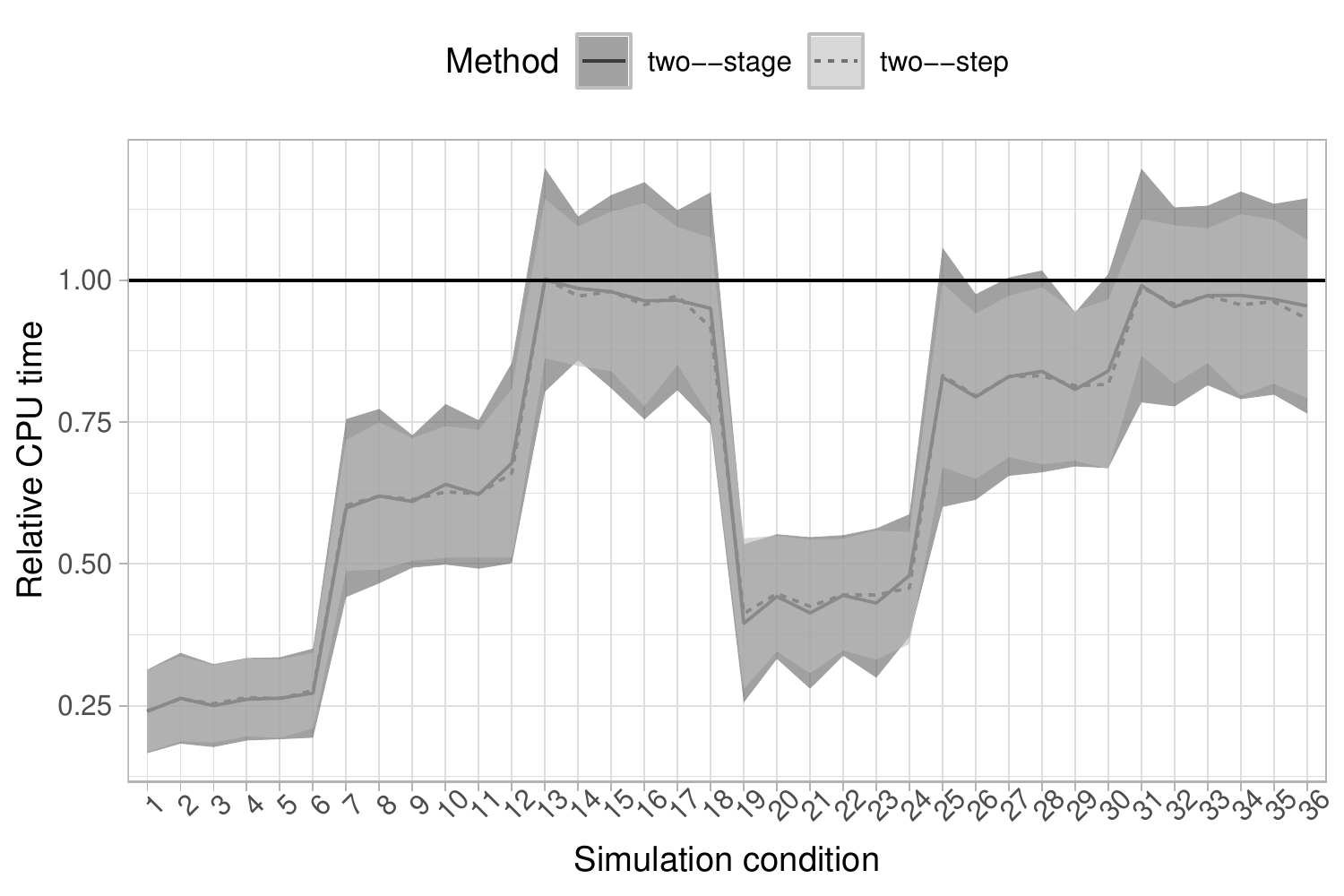}
    \caption{Relative computation time for the one--step, two--stage and two--step estimators for the 24 simulation condition, averaged over the 500 replicates. The one--step estimator's estimation time is taken as reference. Confidence bands based on average values +/- their Monte Carlo standard deviation.}
    \label{fig:cputime}
\end{figure}

The different estimators thus perform essentially identically. Where they differ from each other is in their computational demands. Considering the computation time relative to the simultaneous estimator (Figure \ref{fig:cputime}), we find that both stepwise estimators are always (and up to four times) faster than the simultaneous estimator, and the two--step estimator achieves this with one fewer step compared to the existing two--stage competitor.

% \begin{table}
%     \centering
%     \begin{tabular}{l c c c c}
%     \hline \hline
% Condition		& one--step		&	two--step unc	&	two--step cor	&	two--stage	\\
% \cmidrule{2-5}
% 1    & 0.978  & 0.981  & 1.004  & 0.981 \\ 
% 2    & 1.035  & 1.035  & 1.057  & 1.035 \\ 
% 3    & 1.004  & 1.005  & 1.028  & 1.005 \\ 
% 4    & 1.002  & 1.001  & 1.022  & 1.001 \\ 
% 5    & 0.984  & 0.985  & 1.006  & 0.985 \\ 
% 6    & 1.017  & 1.016  & 1.038  & 1.016 \\ 
% 7    & 1.008  & 1.008  & 1.013  & 1.008 \\ 
% 8    & 1.004  & 1.004  & 1.006  & 1.004 \\ 
% 9    & 1.014  & 1.014  & 1.017  & 1.014 \\ 
% 10    & 0.975  & 0.975  & 0.977  & 0.975 \\ 
% 11    & 1.005  & 1.005  & 1.007  & 1.005 \\ 
% 12    & 1.016  & 1.016  & 1.017  & 1.016 \\ 
% 13    & 1.026  & 1.027  & 1.049  & 1.027 \\ 
% 14    & 0.993  & 0.992  & 1.011  & 0.992 \\ 
% 15    & 0.995  & 0.996  & 1.016  & 0.996 \\ 
% 16    & 1.013  & 1.013  & 1.031  & 1.013 \\ 
% 17    & 1.045  & 1.046  & 1.066  & 1.046 \\ 
% 18    & 1.01  & 1.01  & 1.028  & 1.01 \\ 
% 19    & 1.028  & 1.029  & 1.034  & 1.029 \\ 
% 20    & 1.009  & 1.009  & 1.011  & 1.009 \\ 
% 21    & 1.039  & 1.039  & 1.042  & 1.039 \\ 
% 22    & 1.025  & 1.025  & 1.026  & 1.025 \\ 
% 23    & 1.014  & 1.015  & 1.017  & 1.015 \\ 
% 24    & 1.028  & 1.028  & 1.029  & 1.028 \\ 
% \hline
% \hline
%     \end{tabular}
%     \caption{Correctness of standard error estimation -- computed as standard error over standard deviation -- for the one--step, two--step estimator (uncorrected and corrected SEs) and two--stage estimator averaged over covariate effects.}
%     \label{tab:res_sesd}
% \end{table}

\FloatBarrier

\section {Analysis of cross--national citizenship norms with multilevel LCA}\label{sec:realdata}

In this empirical example, we analyze  citizenship norms in a diverse set of countries. The data are taken from 
the International Civic and Citizenship Education Study (ICCS) 
conducted by the International Association for the Evaluation of Educational Achievement (IEA).
Prior research has used LCA to analyze the first two waves of this survey, which were conducted in 1999 and 2009, to investigate distinctive types of citizenship norms \citep{hooghe2015,hooghe2016,oser2013}.
We focus on the most recent round of the survey, from 2016 \citep{kohler2018}.
The data are from a survey of students in their eighth year of schooling.  
We have data from between 1300 and 7000 respondents in each of 24 countries, as shown in Table \ref{tab:country_sampsize}. 

The respondents answered 12 questions (items) on
how important they think different behaviours are for "being a good adult citizen". These behaviours were always obeying the law (labelled \textit{obey} below), taking part in activities promoting human rights (\textit{rights}), participating in activities to benefit people in the local community (\textit{local}), working hard (\textit{work}), taking part in activities to protect the environment (\textit{envir}), voting in every national election (\textit{vote}), learning about the country’s history (\textit{history}), showing respect for government representatives (\textit{respect}), following political issues in the newspaper,  on the radio, on TV or on the Internet (\textit{news}), participating in peaceful protests against laws believed to be unjust (\textit{protest}), engaging in political discussions (\textit{discuss}), and joining a political party (\textit{party}).

We treat these twelve items as indicators of the individuals' perceptions of the duties of a citizen (\emph{citizenship norms}). The data have a multilevel structure, with individuals as the low-level units and countries as the high-level units. As predictors of low-level latent class membership, we include the respondent's gender, socio-economic status operationalised by the proxy measure of the number of books in their home, and  measures of the respondent's educational expectations, parental education, and if she/he is a non-native language speaker. For details on data cleaning and recoding, see \cite{Oser2023-gt}. 

To compare with previous work on the same data, we fit a multilevel LC model with $T=4$ low-level classes (of individuals within countries) and $M=3$ high-level classes (of countries). 
The same data set was analyzed in \cite{jrss2021} with a multilevel LC model with random intercepts, estimated with a two-stage estimator. We extend \cite{jrss2021}'s model specification by allowing for both random intercepts and random slopes, and we fit the model with the proposed two-step estimator. As the two-step estimator has been shown to be computationally more efficient than the two-stage estimator though with equal performances, for the comparison we include the benchmark simultaneous estimator only.

\begin{table}[!h]
\centering
\begin{tabular}{l c}
\hline
\hline
Country & sample size\\
\cline{1-2}
Belgium  &  2750  \\ % BFL
Bulgaria  &  2682  \\ % BGR
Chile  &  4753  \\ % CHL
Colombia  &  4992  \\ % COL
Denmark  &  5692  \\ % DNK
Germany  &  1313  \\ % DNW
Dominican Republic  &  2779  \\ % DOM
Estonia  &  2770  \\ % EST
Finland  &  3037  \\ % FIN
Hong Kong  &  2553  \\ % HKG
Croatia  &  3655  \\ % HRV
Italy  &  3274  \\ % ITA
Republic of Korea  &  2557  \\ % KOR
Lithuania  &  3422  \\ % LTU
Latvia  &  3000  \\ % LTV
Mexico  &  4987  \\ % MEX
Malta  &  3317  \\ % MLT
Netherlands  &  2692  \\ % NLD 
Norway  &  5740  \\ % NOR
Peru  &  4713  \\ % PER
Russia  &  7049  \\ % RUS
Slovenia  &  2664  \\ % SVN
Sweden  &  2828  \\ % SWE
Taiwan  &  3904  \\ % TWN
\hline \hline
\end{tabular}
\caption{Number of respondents per country of the third wave (2016) of the IEA survey used for the analysis.}
\label{tab:country_sampsize}
\end{table}

\begin{table}
\centering
\begin{tabular}{l c}
\hline
\hline
 & Value \\
log--likelihood & -459295.5\\
 BIC       & 919262.1 \\
 BIC ($J$) & 918778.5\\
 entrR$^2_\text{low}$ & 0.999\\
 entrR$^2_\text{high}$ & 0.999\\
npar & 59\\
\hline \hline
\end{tabular}
\caption{Summary statistics for the measurement model.}
\label{tab:mod_summ}
\end{table}

%\FloatBarrier

\begin{figure}
	\centering
	\includegraphics[width=\textwidth]{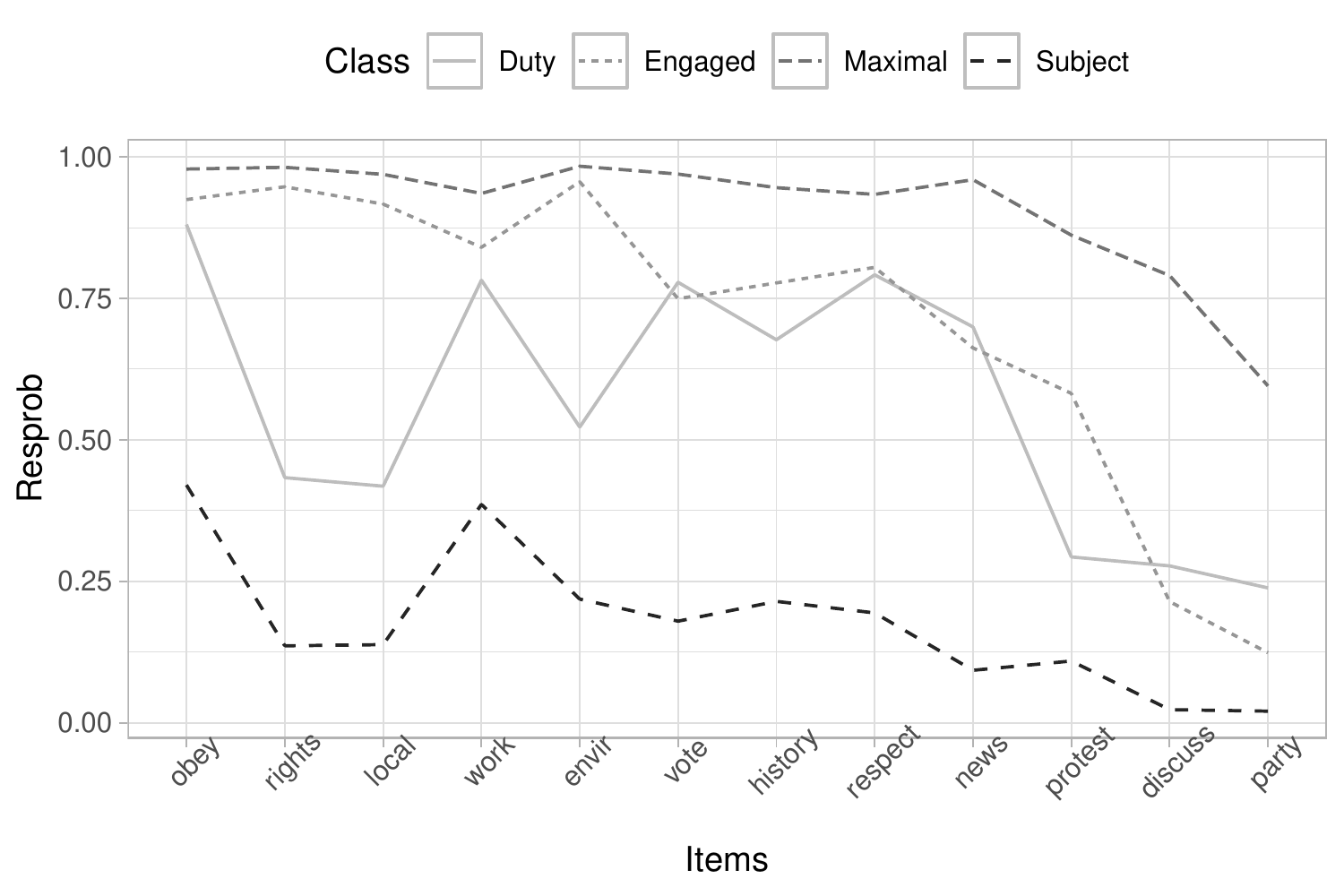}
	\caption{Measurement model at the lower (individual) level: line graph of the class--conditional response probabilities.}
	\label{fig:Figbasic}
\end{figure}

The measurement model, at both levels, presents very well separated classes (Table \ref{tab:mod_summ}).
At the lower level, 
the four latent classes are characterised by their the conditional response probability patterns, as shown in Figure \ref{fig:Figbasic}. 
Two classes present response configurations relating to two relevant and well-known notions of citizenship norms. 
First, a ``Duty" group,  which places high importance on the act of voting, discussing politics, and party activity, while manifesting relatively low interest in protecting human rights and activities to assist the local community.
Second, an ``Engaged" group, which displays higher emphasis on engaged attitudes such as protecting the environment, and lower importance on more traditional citizenship activity items such as membership of political parties. 
In addition, we observe two classes with consistently high and consistently low probabilities of assigning importance to all of the behaviours, here labelled the ``Maximal" and the ``Subject" classes respectively.

\FloatBarrier

At the higher level, the estimated model includes three latent classes for the countries, labelled below as HL1, HL2 and HL3. Considering first the conditional probabilities for the four individual-level classes given these country-level classes (see Table \ref{tab:cond_classproportions}), we can see that HL1 has  clearly the highest conditional probability for the individual ``Duty'' class, HL2 for the ``Maximal'' class and HL3 for the ``Engaged''. The country classes do not differ in probabilities of the passive ``Subject'' class of individuals, which are in any case consistently low. Table \ref{tab:country_clus} shows the assignment of countries to the classes, when the assignment is done based on highest posterior probabilities given the survey responses in the countries. 
Here there are no very clear patterns. Only two countries (Denmark and Netherlands) are assigned to HL1, while the other two classes each include a fairly heterogeneous subset of the rest of the countries. 

%and \ref{tab:cond_classproportions}), the measurement model output can be summarized as follows
%\begin{itemize}
%    \item HL1 includes two advanced democracies, with the highest proportion of ``Duty", and the second highest proportion for "Engaged" ;
%    \item HL2 includes relatively less advanced democracies, with the smallest percentage of ``Engaged" responses, and the highest proportion of ``Maximal"; 
%    \item HL3 is the most heterogeneous group of countries, sharing the highest proportion of ``Engaged" responses. 
%\end{itemize}

\begin{table}
    \centering
    \begin{tabular}{l c c c c}
    \hline \hline
&HL 1 & HL 2 & HL 3 \\ 
\cmidrule{2-4}
Maximal & 0.207 &  0.576 &  0.317 \\ 
Engaged & 0.290 &  0.277 &  0.478 \\ 
Subject & 0.031 &  0.029 &  0.044 \\ 
Duty & 0.471 &  0.118 &  0.161 \\ 
\hline
\hline
    \end{tabular}
    \caption{Estimated proportions of low-level (individual-level) classes conditional on high-level (country-level) class membership.}
    \label{tab:cond_classproportions}
\end{table}

\begin{table}%[!h]
\centering
\begin{tabular}{l c c c}
\hline
\hline
Country & HL 1 & HL 2 & HL 3\\
%\cline{2-4}
\midrule
Belgium & 0 &  0 &  1 \\ 
Bulgaria & 0 &  0 &  1 \\ 
Chile & 0 &  0 &  1 \\ 
Colombia & 0 &  0 &  1 \\ 
Denmark & 1 &  0 &  0 \\ 
Germany & 0 &  0 &  1 \\ 
Dominican Republic & 0 &  1 &  0 \\ 
Estonia & 0 &  0 &  1 \\ 
Finland & 0 &  0 &  1 \\ 
Hong Kong & 0 &  1 &  0 \\ 
Croatia & 0 &  1 &  0 \\ 
Italy & 0 &  1 &  0 \\ 
Republic of Korea & 0 &  1 &  0 \\ 
Lithuania & 0 &  0 &  1 \\ 
Latvia & 0 &  0 &  1 \\ 
Mexico & 0 &  1 &  0 \\ 
Malta & 0 &  0 &  1 \\ 
Netherlands & 1 &  0 &  0 \\ 
Norway & 0 &  0 &  1 \\ 
Peru & 0 &  1 &  0 \\ 
Russia & 0 &  1 &  0 \\ 
Slovenia & 0 &  0 &  1 \\ 
Sweden & 0 &  0 &  1 \\ 
Taiwan & 0 &  1 &  0 \\ 
\hline \hline
\end{tabular}
\caption{Assignment of countries to the high--level classes, based on the maximum a posteriori (MAP) classification rule. $M=3$.}
\label{tab:country_clus}
\end{table}

%\FloatBarrier

Table \ref{tab:regtable} presents estimates of the parameters of main interest in the analysis, the coefficients of the structural model for the lower-level classes given individual-level covariates, separately within each of the higher-level classes. We note first that the one-step and two-step estimates and their standard errors are very similar, as would be expected given the previous simulation results. 

Considering the coefficients themselves, note that they compare each of the other classes to the ``Maximal'' class for whom all of the behaviours are to a greater or less extent considered important to good citizenship. Compared to this class, the relative probability of the (overall quite small) ``Subject'' class for whom none of the behaviours are important, is higher for individuals who are boys, speak the native language at home, have fewer books at home, and have low educational aspirations. The probabilities of the ``Engaged'' class, who are partly similar to ``Maximal'' but place less importance on many of the traditional political activities, are relatively higher for girls, those who have larger number of books at home, and for native speakers. For the ``Duty'' class, which differs from the ``Engaged'' in placing much less importance on direct activism, the probabilities relative to ``Maximal'' are higher for boys and those with low educational aspirations. For the comparisons of other pairs of classes, these estimates also imply, for example, that the probabilities of ``Engaged'' relative to ``Duty'' are generally higher for girls than for boys.
These patterns of the coefficients are broadly similar in each of the country classes, with some variation in detail.

\begin{table}[!h]
\resizebox{!}{.35\paperheight}{
% \resizebox{0.6\textwidth}{!}{
\begin{tabular}{l cc cc cc}
 \hline \hline
HL 1				&	\multicolumn{2}{c}{Engaged}	&	\multicolumn{2}{c}{Subject}	&	\multicolumn{2}{c}{Duty}\\
			%\cline{2-3} 			\cline{4-5} 			\cline{6-7}
		&one--step	&	two--step	&		one--step	&	two--step	&	one--step	&	two--step	\\
		\cline{2-7}
%		& \multicolumn{6}{c}{HL 1} \\
intercept &  0.875*** &  0.944*** &  0.923*** &  0.757*** &  0.945*** &  0.934***\\ 
  & (0.009)  &   (0.009)  &   (0.010)  &   (0.010)  &   (0.159)  &   (0.156)  \\ 
Female &  0.359*** &  0.338*** &  -0.983*** &  -1.072*** &  0.140 &  0.106\\ 
  & (0.092)  &   (0.090)  &   (0.053)  &   (0.052)  &   (0.082)  &   (0.080)  \\ 
Number of books &  -0.016 &  -0.014 &  -0.36*** &  -0.345*** &  -0.166 &  -0.173\\ 
  & (0.080)  &   (0.079)  &   (0.080)  &   (0.079)  &   (0.175)  &   (0.171)  \\ 
Education goal &  0.018 &  0.013 &  -0.819*** &  -0.865*** &  0.232** &  0.207\\ 
  & (0.212)  &   (0.228)  &   (0.181)  &   (0.202)  &   (0.088)  &   (0.095)  \\ 
Mother education &  -0.308** &  -0.311 &  -0.314 &  -0.327 &  -0.007 &  -0.002\\ 
  & (0.116)  &   (0.124)  &   (0.135)  &   (0.148)  &   (0.133)  &   (0.143)  \\ 
Father education &  -0.108 &  -0.117 &  -0.143 &  -0.131 &  -0.164 &  -0.164\\ 
  & (0.256)  &   (0.294)  &   (0.134)  &   (0.134)  &   (0.073)  &   (0.072)  \\ 
Non--native language level &  -0.437*** &  -0.428*** &  -0.03 &  -0.155 &  -0.446*** &  -0.408***\\ 
  & (0.042)  &   (0.042)  &   (0.068)  &   (0.067)  &   (0.065)  &   (0.065)  \\ 

 \midrule
HL 2			&	\multicolumn{2}{c}{Engaged}	&	\multicolumn{2}{c}{Subject}	&	\multicolumn{2}{c}{Duty}\\
			%\cline{2-3} 			\cline{4-5} 			\cline{6-7}
		&one--step	&	two--step	&		one--step	&	two--step	&	one--step	&	two--step	\\
		\cline{2-7}
%		& \multicolumn{6}{c}{HL 2} \\

 intercept &  -0.760*** &  -0.749*** &  -1.404*** &  -1.503*** &  -1.076*** &  -1.099***\\ 
  & (0.064)  &   (0.064)  &   (0.140)  &   (0.139)  &   (0.072)  &   (0.073)  \\ 
Female &  0.199*** &  0.180*** &  -0.651*** &  -0.672*** &  -0.255*** &  -0.278***\\ 
  & (0.036)  &   (0.036)  &   (0.023)  &   (0.023)  &   (0.036)  &   (0.036)  \\ 
Number of books &  -0.133*** &  -0.130*** &  -0.247*** &  -0.265*** &  -0.090 &  -0.087\\ 
  & (0.029)  &   (0.029)  &   (0.029)  &   (0.029)  &   (0.072)  &   (0.071)  \\ 
Education goal &  0.025 &  0.014 &  -0.536*** &  -0.555*** &  -0.306*** &  -0.313***\\ 
  & (0.105)  &   (0.111)  &   (0.079)  &   (0.084)  &   (0.042)  &   (0.045)  \\ 
Mother education &  0.030 &  0.035 &  0.090 &  0.088 &  0.191** &  0.188**\\ 
  & (0.056)  &   (0.059)  &   (0.060)  &   (0.064)  &   (0.060)  &   (0.064)  \\ 
Father education &  0.018 &  0.016 &  -0.160 &  -0.166 &  0.022 &  0.018\\ 
  & (0.157)  &   (0.166)  &   (0.078)  &   (0.079)  &   (0.045)  &   (0.045)  \\ 
Non--native language level &  -0.127*** &  -0.114*** &  -0.306*** &  -0.338*** &  0.299*** &  0.290***\\ 
  & (0.027)  &   (0.027)  &   (0.040)  &   (0.040)  &   (0.037)  &   (0.037)  \\ 

\midrule
HL 3			&	\multicolumn{2}{c}{Engaged}	&	\multicolumn{2}{c}{Subject}	&	\multicolumn{2}{c}{Duty}\\
			%\cline{2-3} 			\cline{4-5} 			\cline{6-7}
		& one--step	&	two--step	&		one--step	&	two--step	&	one--step	&	two--step	\\
		\cline{2-7}
	%	& \multicolumn{6}{c}{HL 3} \\ 
intercept &  0.218*** &  0.260*** &  -0.044 &  -0.217** &  -0.040 &  -0.019\\ 
  & (0.037)  &   (0.037)  &   (0.076)  &   (0.077)  &   (0.071)  &   (0.072)  \\ 
Female &  0.301*** &  0.282*** &  -0.587*** &  -0.616*** &  -0.230*** &  -0.261***\\ 
  & (0.032)  &   (0.032)  &   (0.019)  &   (0.019)  &   (0.035)  &   (0.034)  \\ 
Number of books &  -0.083** &  -0.081** &  -0.358*** &  -0.374*** &  -0.083 &  -0.094\\ 
  & (0.027)  &   (0.027)  &   (0.027)  &   (0.026)  &   (0.059)  &   (0.058)  \\ 
Education goal &  0.148 &  0.124 &  -0.547*** &  -0.544*** &  -0.411*** &  -0.434***\\ 
  & (0.099)  &   (0.106)  &   (0.063)  &   (0.067)  &   (0.035)  &   (0.037)  \\ 
Mother education &  0.040 &  0.044 &  -0.033 &  -0.033 &  0.183*** &  0.176***\\ 
  & (0.050)  &   (0.053)  &   (0.048)  &   (0.051)  &   (0.048)  &   (0.051)  \\ 
Father education &  -0.097 &  -0.097 &  -0.125 &  -0.125 &  0.037 &  0.038\\ 
  & (0.099)  &   (0.106)  &   (0.078)  &   (0.079)  &   (0.040)  &   (0.041)  \\ 
Non--native language level &  -0.426*** &  -0.414*** &  -0.107** &  -0.095 &  -0.006 &  0.004\\ 
  & (0.023)  &   (0.023)  &   (0.039)  &   (0.039)  &   (0.036)  &   (0.036)  \\ 
 \hline \hline
 \end{tabular}}
 \caption{\footnotesize Estimated coefficients of structural models, i.e.\ multinomial logistic models for membership of the four individual-level latent classes conditional on covariates, separately within each of the three country-level latent classes (HL1, HL2 and HL3). The ``Maximal'' class is taken as the reference level for the response class. The number of books available in the respondent's home is treated as a proxy for the respondent's socio--economic status. Both simultaneous (one-step) and the proposed two-step estimators of the same parameters are shown, with standard errors in parentheses.\\ *** $p$--value$<$0.01, ** $p$--value$<$0.05, * $p$--value$<$0.1.}
 \label{tab:regtable}
 \end{table}

\FloatBarrier

Finally, we report CPU time of estimation and the number of iterations until convergence for the two approaches (Table \ref{tab:emp_cputime}). 
In this real-data example, the two-step estimator takes only about 22 seconds to reach convergence, with 26 EM iterations. 
The one-step estimator requires 261 iterations and a running time of around 4.5 minutes to reach convergence. 
Each iteration requires about 0.93 seconds to run for the one-step estimator, while the two-step estimator uses 0.85 seconds and much fewer EM iterations overall.

\begin{table}[!h]
    \centering
    \begin{tabular}{l c c}
    \hline \hline
& CPU time (in seconds) & Number of iterations until convergence \\ 
\cmidrule{2-3}
one--step & 242.89 & 261  \\ 
two--step & 22.01  & 26 \\
\hline
\hline
    \end{tabular}
    \caption{CPU time to estimation in seconds, and number of iterations until convergence for the two methods - one--step and two--step estimators.}
    \label{tab:emp_cputime}
\end{table}

\FloatBarrier
\section{Discussion}

In this paper we proposed a two--step estimator for the multilevel latent class model with covariates. It concentrates the estimation of the measurement model in a single first step. In the second step, covariates are added to the model, keeping the measurement model parameters fixed. 
The approach represents a simplification over the recently proposed two-stage estimator \citep{bakk2021sem} by having only two steps instead of multiple sub-steps in estimating the measurement model. 

We discussed model identification of the unconditional model, derived an Expectation Maximization algorithm for efficient estimation of both steps and presented second-step asymptotic standard errors that account for the variability in the first step.
The simplified two-step procedure makes it possible to apply the standard theory of \cite{gong1981pseudo} for obtaining unbiased standard errors, a further improvement over the two-stage estimator.
An effective initialization strategy, using (dissimilarity--based) cluster analysis, was also proposed.

In the simulation study, we observed that the performance of the proposed estimator in terms of bias is very similar to the benchmark simultaneous (full-information ML) estimator --- and similar to that of the two-stage estimator --- with nearly no efficiency loss. The two-step estimator was up to 4 times faster than the simultaneous estimator. 
It should be mentioned that, in conditions where the entropy of the LC model is low, all estimators show relatively higher variability and bias,
a finding in line with previous research on stepwise  estimators for single--level LC models \citep{vermunt:10}.

%This with one less step compared to the already existing two--stage competitor. 
In the real data example, we found interesting lower and higher level class configurations, consistent with existing literature on the topic of citizenship norms (see, e.g., \citealp{oser2022}). In the structural model, 
the model allows us to investigate the associations between covariates
and the latent classes, including the possibility of 
group-level heterogeneous effects of covariates on lower class membership.
In addition, we found a considerable CPU running time difference between the one--step and the two--step estimators, which was even larger than what we observed in the more controlled simulation environment. 
More specifically, whereas the former required 4.5 minutes to reach convergence, the latter only needed 22 seconds. 
From an applied user's perspective, such a CPU time gain can be substantial on a larger scale. 
As an example, consider a data set with larger low- and high- level sample sizes: if simultaneous estimation took 2 hours, our two-step estimator would produce final estimates in only roughly 12 minutes. 
We expect, based on existing literature on two--step estimators (see, e.g., \citealp{dimari2021}), such a gap to increase in model complexity - i.e. number of lower/higher level classes and/or available predictors. The difference in time is also multiplied if the models are estimated repeatedly, for example when different sets of covariates or different numbers of latent classes are explored.

There are some issues that deserve future research. 
First, while we describe two possible approaches for class selection in Section \ref{sec:classel}, this is not the main focus of the current work. 
Further research should investigate class selection using the different estimators.
Second, we have proposed estimates for the asymptotic variance--covariance matrix based on the outer product of the score. Deriving Hessian-- and/or sandwich--based \citep{white1982maximum} standard errors, e.g. for small high--level sample size and complex sampling scenarios, can be interesting topics for future work. 
Third, we have discussed multimodality of the likelihood surface as a long--standing well--known characteristic feature related, in general, to mixture models.
The EM algorithm's properties have been largely studied over the years - i.e., monotonicity, and global convergence (see, e.g., \citealp{redner1984}).
The EM has several advantages, e.g., low cost per iteration, economy of storage and
ease of programming.
However, in practice, due to multimodality, convergence to global or local optima depends on the choice of the starting point \citep{wu1983}.
As such, there is no systematic, neither theoretical nor simulation based, study of the behavior of the EM with two--step estimators.
We speculate that, given that the second step operates in a lower dimensional space compared to simultaneous estimation, two--step estimators should somewhat restrain the initialization problem. 
This point, being not the focus of the current work, certainly deserves specialized attention. 
For this, and related matters, we defer to future research.

\bibliography{zsuzsa2jk}
\bibliographystyle{apacite}

\section*{Acknowledgements}
The Authors thank Johan Lyrvall for his valuable comments on the manuscript.
Di Mari acknowledges financial support from a University of Catania grant (Starting Grant FIRE, PIACERI 2020/2022), and Oser by a European Union grant (ERC, PRD, project number 101077659). 

\appendix

\section{Computation of the score vector for the multilevel latent class model}
\subsection{The unconditional multilevel LC (first step)}
Let us reparametrize the unconditional multilevel LC model of Equation \eqref{eq:uncondMLC} according to the following log-linear equations

\begin{equation}
  \label{eq:paramMLC}
\begin{cases}
\log\left[\displaystyle \frac{\phi_{h \vert t}}{1-\phi_{h \vert t}} \right] = \beta_{h \vert t} \\
\log\left[\displaystyle\frac{\omega_m}{\omega_1}\right] = \alpha_{m} \\
\log\left[\displaystyle\frac{\pi_{t \vert m}}{\pi_{1 \vert m}}\right] = \gamma_{t \vert m},
\end{cases}
\end{equation}

In addition, let us conveniently rewrite \eqref{eq:Ecdll_MLCA} as follows

\begin{equation}\label{eq:Qfun}
    Q(\balpha,\bGamma,\bB) = Q(\balpha) + Q(\bGamma) + Q(\bB),
\end{equation}

where $\balpha = (\alpha_2,\dots,\alpha_M)^{\prime}$, $\bGamma$ is a $T-1 \times M$ matrix with elements $\gamma_{t\vert m}$, for $m=1,\dots,M$ and $t=2,\dots,T$,  $\bB$ is an $H \times T$ matrix with elements $\beta_{h\vert t}$ for $t=1,\dots,T$ and $h=1,\dots,H$, and

\begin{align}
Q(\balpha) &= \sum_{j=1}^J \sum_{m=1}^M  \widehat{u}_{j,m} \log(\omega_m) \label{eq:Qfunext1}\\
Q(\bGamma) &= \sum_{j=1}^J \sum_{i=1}^{n_j} \sum_{m=1}^M \sum_{t=1}^{T} \widehat{v}_{i,j,t,m} \log(\pi_{t \vert m}) \label{eq:Qfunext2}\\
Q(\bB) &= \sum_{j=1}^J \sum_{i=1}^{n_j} \sum_{m=1}^M \sum_{t=1}^{T} \widehat{v}_{i,j,t,m}\{ Y_{ijh}\log(\phi_{h \vert t}) + [1-Y_{ijh}]\log(1-\phi_{h \vert t}) \label{eq:Qfunext3} \}.
\end{align}

Recalling that $\partial \ell (\param) / \partial \param^{\prime} = \partial Q / \partial \param^{\prime} $, the $ij$-th contribution to the score has the following three blocks, with generic elements

\begin{align}
\bs_{ij,\param_1}(\widehat{\alpha}_m)  &= \widehat{u}_{j,m} - \omega_m \label{eq:score1}\\
\bs_{ij,\param_1}(\widehat{ \gamma}_{t \vert m}) &= (\widehat{q}_{i,j,t\vert m} - \pi_{t\vert m})\widehat{u}_{j,m} \label{eq:score2}\\
\bs_{ij,\param_1}(\widehat{ \beta}_{h \vert t}) &= \sum_{m=1}^M  \widehat{v}_{i,j,t,m}(Y_{ijh} - \phi_{h \vert t}) \label{eq:score3}.
\end{align}

Thus, an estimate of $\bSigma_{11}$ can be obtained as follows

\begin{equation}\label{eq:sig11est}
\widehat{\bSigma}_{11} = N^{-1} \sum_{j=1}^J \sum_{i=1}^{n_j} \bs_{ij}(\widehat{\balpha},\widehat{\bGamma},\widehat{\bB}) \text{ } \bs_{ij}(\widehat{\balpha},\widehat{\bGamma},\widehat{\bB})^{\prime}   
\end{equation}

\subsection{The multilevel LC model with covariates (second step)}

Let us define $\pi^{ij}_{t \vert m} = \frac{\exp(\bgamma_{tm}^{\prime} \mathbf{Z}_{ij})}{1+\sum_{s=2}^{T}\exp(\bgamma_{tm}^{\prime} \mathbf{Z}_{ij})}$. The $Q$ function of Equation \eqref{eq:cdll_MLCcov} can be rewritten under the log--linear parametrizations introduced above, except for the second block which is as follows

\begin{equation}\label{eq:score2cov}
   Q(\bGamma) = \sum_{j=1}^J \sum_{i=1}^{n_j} \sum_{m=1}^M \sum_{t=1}^{T} \widehat{v}_{i,j,t,m} \log(\pi^{ij}_{t \vert m}) 
\end{equation}

The second block of the $ij$-th contribution to the score as generic $K+1$ contributions
\begin{equation}
    \bs_{ij,\param_2}(\widehat{\bgamma}_{tm})  = 
\widehat{u}_{j,m}(\widehat{q}_{i,j,t\vert m} - \pi^{ij}_{t \vert m})\bz_{ij}.
\end{equation}
% (mU.col(k) - w_i.col(k)).t() *(mWei % X)
% mU is cPX and mWei is mPW

% \section{Useful notes for partial derivatives}

% \begin{equation}\label{eq:condMLC}
% \mathcal{L}_j = \omega_1 \sum_{m=2}^M \exp(\alpha_m) \prod_{i=1}^{n_j} \pi_{1 \vert m} \sum_{t=2}^{T} \exp(\gamma_{t \vert m}) \prod_{h=1}^{H} \overline{\phi}_{h \vert t} \exp(Y_{ijh}\beta_{h \vert t}),
% \end{equation} 

% or

% \begin{equation}\label{eq:condMLC}
% \mathcal{L}_{ij} = \omega_1 \sum_{m=2}^M \exp(\alpha_m) \pi_{1 \vert m} \sum_{t=1}^{T} \pi_{t \vert m} \prod_{h=1}^{H} \overline{\phi}_{h \vert t} \exp(Y_{ijh}\beta_{h \vert t}),
% \end{equation} 

% where 

% \begin{equation}
%   \omega_1  = \frac{1}{1+\sum_{l=2}^M \exp(\alpha_m})\text{ ,}\quad
%   \pi_{1 \vert m} = \frac{1}{1+\sum_{s=2}^T \exp(\gamma_{s \vert m})}\text{ ,}\quad
%   \overline{\phi}_{h \vert t} = \frac{1}{1 + \exp(\beta_{h \vert h})}.
% \end{equation}

% \begin{equation}
%   \frac{\partial \omega_1}{\partial \alpha_l}  = -\frac{\exp(\alpha_l)}{[1+\sum_{l=2}^M \exp(\alpha_l)]^2}\text{ ,}\quad
%   \frac{\partial \pi_{1 \vert m}}{\partial \gamma_{r \vert m}} = -\frac{\exp(\gamma_{r \vert m})}{[1+\sum_{s=2}^T \exp(\gamma_{s \vert m})]^2}\text{ ,}\quad
%   \frac{\partial \overline{\phi}_{h \vert t}}{\partial \beta_{k \vert t}} = -\frac{\exp(\beta_{k \vert t})}{[1 + \exp(\beta_{k \vert t})]^2}.
% \end{equation}

\section{Extra tables and figures}

\begin{table}
    \centering
    \begin{tabular}{l c c}
    \hline \hline
Condition		& two--stage	&	two--step\\
\cmidrule{2-3}
1    & 0.985  & 0.985 \\ 
2    & 1.000  & 1.000 \\ 
3    & 0.99  & 0.99 \\ 
4    & 0.995  & 0.995 \\ 
5    & 0.989  & 0.989 \\ 
6    & 0.998  & 0.998 \\ 
7    & 0.997  & 0.997 \\ 
8    & 1.000  & 1.000 \\ 
9    & 0.997  & 0.997 \\ 
10    & 0.999  & 0.999 \\ 
11    & 0.999  & 0.999 \\ 
12    & 1.000  & 1.000 \\ 
13    & 1.000  & 1.000 \\ 
14    & 1.000  & 1.000 \\ 
15    & 1.000  & 1.000 \\ 
16    & 1.000  & 1.000 \\ 
17    & 1.000  & 1.000 \\ 
18    & 1.000  & 1.000 \\ 
19    & 0.983  & 0.983 \\ 
20    & 0.998  & 0.998 \\ 
21    & 0.993  & 0.993 \\ 
22    & 1.005  & 1.005 \\ 
23    & 0.996  & 0.996 \\ 
24    & 1.004  & 1.004 \\ 
25    & 1.001  & 1.001 \\ 
26    & 0.999  & 0.999 \\ 
27    & 0.997  & 0.997 \\ 
28    & 1.000  & 1.000 \\ 
29    & 1.001  & 1.001 \\ 
30    & 0.999  & 0.999 \\ 
31    & 0.999  & 0.999 \\ 
32    & 1.000  & 1.000 \\ 
33    & 1.000  & 1.000 \\ 
34    & 1.000  & 1.000 \\ 
35    & 1.000  & 1.000 \\ 
36    & 1.000  & 1.000 \\ 
\hline
\hline
    \end{tabular}
    \caption{Average relative efficiency for the two--step and two--stage estimator relative to the one-step estimator (SD over benchmark one-step SD), averaged over covariate effects.}
    \label{tab:res_sesd}
\end{table}

\end{document}